\documentclass[
aps, 
prb, 
10pt,  
amssymb, 
amsmath, 
superscriptaddress, 
tightenlines, 
twocolumn, 
notitlepage, 
] {revtex4-2}
\usepackage{graphicx}
\usepackage[colorlinks, linkcolor = blue, citecolor = blue, urlcolor=blue, pdfborder={0 0 0 [0 0]},bookmarks=false]{hyperref}
\usepackage{bm}
\usepackage{physics}
\usepackage{amssymb}
\usepackage{amsmath}
\usepackage[dvipsnames]{xcolor}
\usepackage{soul}

\usepackage{lipsum}

\newcommand{\be}{\begin{equation}}
\newcommand{\ee}{\end{equation}}

\begin{abstract} 
We propose a universal mechanism for the Josephson diode effect in short Josephson junctions. The proposed mechanism is due to finite Cooper pair momentum and is a manifestation of simultaneous breaking of inversion and time-reversal symmetries.  The diode efficiency is up to 40 \%, which corresponds to an asymmetry between the critical currents in  opposite directions  $I_{c+}/I_{c-} \approx 230$ \%. We show that this arises  from both the Doppler shift of the Andreev bound state energies and the phase-independent asymmetric current from the continuum.  Finally, we propose a simple scheme for achieving  finite-momentum pairing, which  does not rely on spin-orbit coupling and thus greatly expands existing platforms for the observation of supercurrent diode effects. 
\end{abstract}

\begin{document}
\author{Margarita Davydova}
\affiliation{Department of Physics, Massachusetts Institute of Technology, Cambridge, MA 02139, USA }
\author{Saranesh Prembabu}
\thanks{Current address: Department of Physics, Harvard University, Cambridge, MA 02138}
\affiliation{Department of Physics, Massachusetts Institute of Technology, Cambridge, MA 02139, USA }

\author{Liang Fu}
\affiliation{Department of Physics, Massachusetts Institute of Technology, Cambridge, MA 02139, USA }

%\date{\today}
\title{Universal Josephson diode effect}

\maketitle

% \par\noindent\rule{\textwidth}{0.4pt}
% \vspace{-30 pt}
% \tableofcontents
%\par\noindent\rule{\textwidth}{0.4pt}

\textit{Introduction. --- }
Recently, there has been a surge of interest in nonreciprocal phenomena in superconductors. In particular, recent experiments have observed an asymmetry between forward and reverse critical currents $I_{c+} \neq I_{c-}$ in superconducting films  \cite{ando2020observation,bauriedl2021supercurrent, shin2021magnetic} and Josephson junctions \cite{bocquillon2017gapless,wu2021realization,baumgartner2021supercurrent,pal2021josephson,baumgartner2021effect,pal2021josephson}.  In the presence of such a nonreciprocity, currents of magnitudes in the range between $I_{{c+}}$ and $I_{c-}$ can flow without resistance in only one direction, resulting in supercurrent diode effect. The dissipationless superconducting diodes should be contrasted with the asymmetric current-voltage characteristics of conventional semiconductor diodes, which are based on resistive and dissipative transport.

The supercurrent diode effect can occur when the free energy is asymmetric under the sign change of the supercurrent. This requires time-reversal symmetry breaking, which can be achieved by an external magnetic field or an exchange field from magnetic proximity effect. As an example, in the presence of Zeeman field, a noncentrosymmetric superconductor with spin-orbit interaction can acquire finite Cooper pair momentum, resulting in the so-called helical superconducting state~\cite{PhysRevLett.87.037004,yuan2021topological}. Recent theory predicts \cite{yuan2021supercurrent,daido2021intrinsic,he2021phenomenological,scammell2021theory,ilic2021effect} that such systems possess supercurrent diode effect, where the  critical currents along and against the direction of the Cooper pair momentum have different magnitudes. Thus, the supercurrent diode effect is a direct manifestation of unconventional superconductivity with broken time-reversal and inversion symmetries.

Supercurrent nonreciprocity has also been observed in a number of experiments on Josephson junctions~\cite{bocquillon2017gapless,wu2021realization,baumgartner2021supercurrent,pal2021josephson,baumgartner2021effect,pal2021josephson,doi:10.1063/1.4984142}, often referred to as Josephson diode effect (JDE). While several theoretical proposals 
based on various physical mechanisms have been put forward \cite{yokoyama2013josephson,yokoyama2014anomalous,zazunov2009anomalous,brunetti2013anomalous,PhysRevB.86.214519,PhysRevB.98.075430,pal2019quantized,kopasov2021geometry,PhysRevLett.99.067004,zhang2021general}, a clear understanding of the microscopic origin of the Josephson diode effect observed in  experiments is still lacking.   %However, the successful theories of supercurrent diode effect in superconducting films \cite{yuan2021supercurrent,daido2021intrinsic,he2021phenomenological}  all involve finite-momentum Cooper pairing. 
Very recently, an experiment~\cite{pal2021josephson} showed simultaneous occurrence of the Josephson diode effect and finite Cooper pair momentum in a superconductor-normal-superconductor junction where two niobium electrodes were coupled by a thin flake of topological semimetal NiTe$_2$. Moreover, the observed features of the Josephson diode effect, such as the temperature and the magnetic field dependence of $\Delta I_c \equiv I_{c+} - I_{c-}$, were accounted for by a phenomenological model based on finite-momentum Cooper pairing~\cite{pal2021josephson}. % formed from spin-helical electrons on the surface of NiTe$_2$. %The connection be is supported by a phenomenological theory. 
This suggests a previously unknown link between the Josephson diode effect and finite-momentum Cooper pairing. However, a microscopic theory relating these two phenomena is yet to be developed.

\begin{figure}[t] 
	\includegraphics[width= 1\columnwidth]{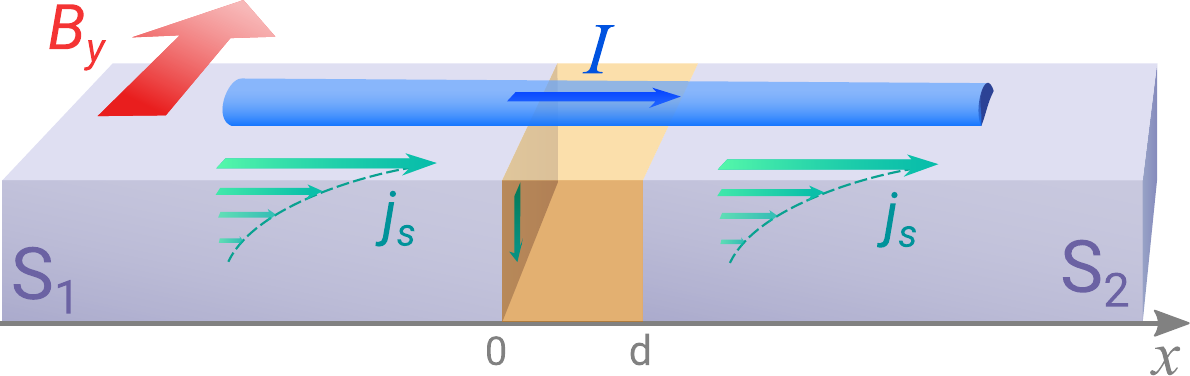}
	\caption{ Schematic illustration of   a short junction  formed by placing a metallic nanowire bridge  on top of two superconducting slabs. Due to an applied in-plane magnetic field $B_y<B_{c1}$,  screening currents $j_s$ emerge in the superconductors leading to finite-momentum Cooper pairing and periodically modulated pairing potentials ($\Delta_{1,2} \sim e^{2 i q x}$) at the surface of the slabs $S_1$ and $S_2$. Consequently, the proximity-induced pairing potential in the metallic nanowire acquires the same spatial modulation.   
	} \label{fig1D1}
\end{figure}
In this work, we present a universal theory of Josephson diode effect due to Cooper pair momentum in short Josephson junctions. An analytical formula for the Josephson current %as a function of on the phase difference in the junction 
is obtained for a short junction between finite-momentum superconductors, which generalizes the well-known result for zero-momentum superconductors. We find an asymmetry between the critical currents in opposite directions that is directly related to the Cooper pair momentum. 
We propose a simple scheme for achieving  finite-momentum pairing and thus inducing the JDE, which is based on the Meissner effect. This scheme only involves a small magnetic field $H<H_{c1}$, does not require spin-orbit coupling, and 
is applicable to {\it all} superconductors. Thus, our work greatly expands the material platform for observing the JDE.    

Our mechanism of the Josephson diode effect has the following origin. The presence of finite Cooper pair momentum  results in Doppler energy shift of  quasiparticle energies by $\pm q v_F$ for left and right movers with momentum close to $\pm k_F$, respectively. Since  left and right movers carry currents in opposite directions, by breaking their degeneracy  the Doppler shift causes the Josephson current to be direction-dependent and thus gives rise to the Josephson diode effect, as we show below. We find that the contribution from the continuum of states plays important role for determining the magnitude of the asymmetry between the critical currents in opposite directions.

%make a theoretical progress in answering these open questions by identifying a universal mechanism for Josephson diode effect in short Josephson junctions. 
%Our explanation is remarkably simple and allows to determine a universal ingredient that is necessary for achieving this effect: it is finite Cooper pair momentum, regardless  how exactly it is achieved (we address several possible ways to do so later in the text) and  the details of the short junction. %This picture clearly shows that, despite being a  distinct effect,  the JDE, just like supercurrent diode effect,  is a robust manifestation of simultaneous inversion- and time-reversal symmetry breaking. 

Unlike previous proposals, the Josephson diode effect we found in short junctions is universal: it is independent of the junction parameters and  occurs already in the ballistic limit.  It arises from and provides a measure of the Doppler shift of quasiparticle energy due to Cooper pair momentum.   
The proposed mechanism does not rely on  scattering between multiple conduction channels~\cite{zazunov2009anomalous,yokoyama2013josephson,yokoyama2014anomalous,brunetti2013anomalous,PhysRevB.86.214519}, layered magnetic structures ~\cite{pal2019quantized}, curved geometry of a nanowire~\cite{kopasov2021geometry}, or doped Mott insulator region at the interface~\cite{PhysRevLett.99.067004}. We find that the diode effect remains present in the presence of disorder in the system and potential barrier in the junction.

\textit{Finite-momentum pairing from the Meissner effect. ---}
%In order to explicate the mechanism for the JDE, 
We consider a short weak link between two superconducting regions as %Josephson junction %made out of single-channel nanowire 
shown in Fig.~\ref{fig1D1}, where both superconductors possess Cooper pair momentum $q$.  Thus,  % and there is an additional phase difference $\varphi$ between them. 
the superconducting order parameter (pair potential) in regions 1 and 2 is 
\begin{eqnarray}
\Delta_1(x) = \Delta e^{2 i q x}, \; \; \Delta_2(x)= \Delta e^{i \varphi + 2 i q x} \label{setup}
\end{eqnarray}
respectively, where $\varphi$ is the overall phase difference between them. 
Note that $q\neq 0$ breaks both inversion and time-reversal symmetry, which is necessary for the Josephson diode effect. 
Our goal is to calculate the current-phase relation for this setup and demonstrate the asymmetry between the critical currents in opposite directions.

While finite-momentum superconductivity is thought to be elusive, the finite-momentum Josephson junction --- the subject of our study --- can be achieved by placing a normal metal bridge or semiconductor wire on top of two conventional superconductors, as shown in Fig.~\ref{fig1D1}. Applying a small in-plane magnetic field $B_y<B_{c1}$ in the direction perpendicular to the wire induces a screening current on the surface of each superconductor. Correspondingly, the superconducting order parameter on the surface develops a spatially modulated phase $\theta(x)=q x$, where $q$ is the Cooper pair momentum (assuming the gauge $A=0$ within the superconductor). For thick superconducting slabs, $q \approx B_y \lambda_L$ where $\lambda_L$ is the London penetration depth. 

By means of superconducting proximity effect, the induced pair potential in regions 1 and 2 of the normal metal or semiconductor bridge (see  Fig.~\ref{fig1D1}) inherits the spatially modulated phase $\theta(x)$ from the superconducting order parameters on the surface of the two superconductors. Our setup thus realizes the required condition (\ref{setup}) for finite-momentum Josephson junction. Here, the proximity-induced gap $\Delta$ is generally smaller than the gap of the parent superconductor. The Cooper pair momentum $q$ is controlled by the external magnetic field.   Note that this scheme for creating finite momentum pairing does not rely on spin-orbit coupling, but instead utilizes surface screening current that is universally present in the Meissner phase of all superconductors. Indeed, a clear evidence of finite-momentum pairing has recently been observed in thin Bi$_2$Te$_3$ layers proximitized by the superconductor NbSe$_2$ under a small magnetic field on the order of $10$ mT~\cite{zhu2021discovery}.  We also discuss alternative ways to achieve finite Cooper pair momentum later.

\textit{The current-phase relation. ---}
%\cite{beenakker1991universal,beenakker1992three}
We find the current-phase relation for a short junction with two superconducting regions at $x<0$ and $x>d$, where the junction length is much smaller than the induced coherence length $d \ll \xi_{ind} \propto \frac{v_F}{\Delta}$.
The  order parameters in the superconducting regions are  given in eq.~\eqref{setup}, and in the normal region $0<x<d$  we assume $\Delta(x) = 0$. Without loss of generality, we assume that $\Delta >0$.  In what follows, we solve for $N=1$ mode in the junction. It is straightforward to generalize this approach to the case of multiple modes. 

 The BdG Hamiltonian in the superconducting regions is
\be \label{eq:H_text1}
\begin{split}
 %   &\mathcal H = \int d x(\psi_{\uparrow}^\dagger, \psi_{\downarrow})H_{BdG}(\psi_{\uparrow}, \psi_{\downarrow}^\dagger)^T\\
     &\mathcal H = \int d x(\psi_{\uparrow}^\dagger, \psi_{\downarrow})
     \begin{pmatrix}
 -\frac{\partial_x^2}{2 m}  - \mu & \Delta (x) \\ 
\Delta^* (x) & \frac{\partial_x^2}{2 m}  + \mu  
\end{pmatrix}
     \left( 
     \begin{matrix}
     \psi_{\uparrow} \\ 
     \psi_{\downarrow}^\dagger
     \end{matrix}
     \right) \\
% &H_{BdG} = 
\end{split}
\ee
where $\Delta(x)$ in left and right superconducting regions are defined in Eq.(\ref{setup}). 
Assuming that the chemical potential is large $\mu \gg \Delta$, we linearize the kinetic energy near momenta $\pm k_F$, which correspond to right and left movers, respectively. We also rotate the wavefunctions in the superconducting regions 1 and 2 according to $e^{i \left ( q x + \varphi_{1,2}/2 \right ) \tau_z}$, where $\varphi_1 = 0$, $\varphi_2 = \varphi$, and $\tau_z$ is the Pauli matrix in the Nambu basis. Thus we arrive at an especially simple  Hamiltonian:
\begin{eqnarray}
&\mathcal{H} = \frac{1}{2} \int d x \left [  (c_{+}^\dagger,c_{\downarrow -})H_{+}
\binom{c_{\uparrow +}}{c_{\downarrow -}^\dagger}+  (c_{\uparrow -}^\dagger,c_{\downarrow +})H_{-}\binom{c_{\uparrow -}}{c_{\downarrow +}^\dagger}\right ] \nonumber\\
&H_+ = \begin{pmatrix}
v_F(-i\partial_x + q) & \Delta\\ 
\Delta & -v_F(-i\partial_x - q) 
\end{pmatrix}, \label{eq:H_text} \\ 
&H_- =\begin{pmatrix}
-v_F(-i\partial_x + q) & \Delta\\ 
\Delta & v_F(-i\partial_x - q) 
\end{pmatrix},\nonumber  
\end{eqnarray} 
where the notation `$+/-$' pertains to the states with momentum near $\pm k_F$, respectively. 
Note that in an infinite system described by this Hamiltonian, 
 due to the Doppler effect from Cooper pair momentum, $\Delta \pm q v_F$, the quasiparticle dispersion possesses two spectral gaps for quasiparticles moving in the right and left directions, correspondingly. 

We use the scattering matrix formalism developed in refs.~\cite{beenakker1991universal,beenakker1992three} in order to find the spectrum of the bound states. 
In the absence of normal reflection, it is given by roots of the equation:
\be \label{main_text}
\left ((r_A^+)^2 - e^{2 i q d + i \varphi} \right) \left ((r_A^-)^2 - e^{-2 i q d - i \varphi} \right) =0.
\ee
The effect of the normal reflection is discussed later in the text. Here, $r_A^{\pm}$ are up to a phase  proportional to the Andreev reflection amplitudes at the interface for right- and left-moving particles:
\begin{align}
r_A^{\pm} = \frac{E \mp v_F q}{\Delta } - i \sqrt{1 - \left (\frac{E\mp v_Fq}{\Delta} \right)^2}.
\end{align}
The presence of two different coefficients $r_A^{\pm}$ is a direct consequence of the broken time-reversal and inversion symmetries. 

\begin{figure}[t] 
	\includegraphics[width= 1\columnwidth]{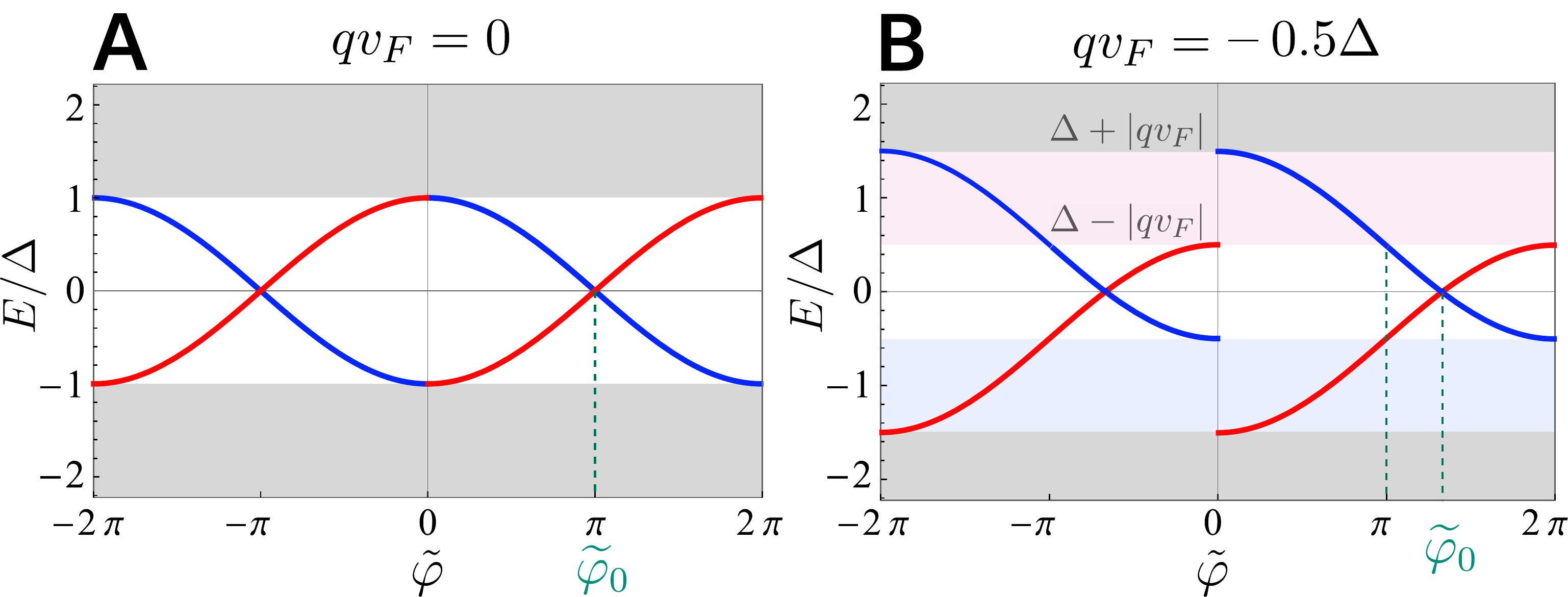}
	\caption{ Spectra of the bound states in the junction at (\textbf{A}) zero and (\textbf{B}) nonzero values of Cooper pair momentum $2q$ in perfectly transparent junction. The bound states originating form left- and right-moving states are shown in  blue and red, respectively.
	} \label{fig:spectrum}
\end{figure}

%We treat positive and `fictitious' negative energies on equal footing here, carefully keeping track of when they lead to double counting. 
In the absence of any normal reflection, the right- and left-moving states are decoupled and produce two separate bound states:
\begin{align} \label{eq:ABS_energy}
  E_1 &=  -\Delta \cos\frac{\tilde \varphi}{2} + v_F q,  \quad -\Delta + qv_F < E < \Delta + qv_F  \nonumber \\
  E_2 &=  \Delta \cos\frac{\tilde \varphi}{2} - v_F q, \quad  -\Delta - qv_F < E < \Delta - qv_F
\end{align}
where we introduced the notation $\tilde \varphi = \varphi + 2 q d$. The constant phase  $2 q d$ is negligible in short junctions and is not relevant to the discussion of the supercurrent nonreciprocity.  

The spectrum is shown in Fig.~\ref{fig:spectrum};  for $\tilde \varphi$ near $0$ and $2\pi$ the branches enter the continuum and the resulting spectrum is $2\pi$-periodic in $\tilde  \varphi$. The bound state energies  generalize the known result for the short junctions at $q=0$~\cite{furusaki1991,beenakker1991universal}. In the presence of finite Cooper pair momentum $q \neq 0$, 
the key difference is that  the branches for right- and left-moving particles are shifted in energy by $\pm q v_F$.

The Josephson current can be determined as $I =  \frac{2 e}{\hbar} \frac{d F}{d \varphi}$, where $F$ is the free energy of the system~\cite{beenakker1992three}.
At zero temperature, the current is:
\be
\begin{split}
I = -  \frac{2e}{\hbar}\sum_{E>0} \frac{dE}{d\varphi} -  \frac{2e}{\hbar} \int     _{E\,  \in \text{ cont.}}^\infty dE \, E  \frac{d \nu (E)}{d \varphi}.
\end {split}
\ee
The first term is the contribution from the bound states and the second term corresponds to the current from the continuum of states; here, $\nu(E)$ is the density of states in the continuum. Thus, we decompose the total current as $I = I_{\text{bound}} + I_{\text{cont}}$. The contribution  from the bound states equals
\be \label{J_bound}
I_{\text{bound}} = - \frac{2 e}{\hbar} \frac{d |E_{\text{bound}}|}{d \varphi} = \frac{ e \Delta}{\hbar} \sin \frac{\tilde \varphi}{2} \text{sgn} \left( \Delta \cos \frac{\tilde \varphi}{2}   - q v_F \right)
\ee
Unlike in conventional short Josephson junctions, the branch change does not occur at  $\pi$ anymore but is determined by the zero of argument of $\text{sgn}(..)$ function, namely  $\widetilde{\varphi}_0 = 2 \arccos\left( \frac{q v_F}{\Delta}\right)$, which is shown in Fig.~\ref{fig:spectrum}B. This is the consequence of the shift of the two bound state energies seen Fig.~\ref{fig:spectrum}B.

Let us now discuss the second contribution to the Josephson current. When time-reversal symmetry is broken, the continuum of states is known to contribute to the Josephson current (see refs. \cite{san2013multiple,kopasov2021geometry} and the discussion in the Supplementary). The density of states $\nu(E)$ in the junction  can be evaluated as~\cite{beenakker1992three}, 
\be \label{dos}
\nu(E)=-\frac{1}{\pi } \Im \frac{\partial}{\partial E} \ln \det\left ( 1- s_A s_N\right)  + \text{const.}
\ee
where $s_A$ and $s_N$ are the scattering matrices transforming the wavefunctions due to Andreev reflection at the interfaces and due to the propagation/scattering in the weak link; the `const.' is the phase-independent part of the density of states. The expression $\det\left ( 1- s_A s_N\right)$ equals left-hand side of eq.~\eqref{main_text}, where in the continuum, the expressions for $r_A^\pm$ have to be properly analytically continued. In the absence of normal reflection, the energy range for the continuum of states formed by decoupled right- and left-movers is $\{ E> \Delta +qv_F, E< -\Delta +q  v_F  \}$ and $\{E> \Delta-qv_F, E< -\Delta-q v_F \}$, respectively. Thus,  the current originating from the continuum is:
\be
\begin{split}
I_{\text{cont}} = &\frac{2e}{\hbar} \frac{1}{\pi} \Im \left [\int_{\Delta -qv_F}^\infty dE \, E \frac{\partial }{\partial E} \frac{d}{d\varphi} \ln \left ( (r_A^-)^2 - e^{ -i \tilde \varphi} \right)+ 
\right.
\\
&+ \left.\int_{\Delta + qv_F}^\infty dE \, E \frac{\partial }{\partial E} \frac{d}{d\varphi} \ln \left ( (r_A^+)^2 - e^{ i \tilde \varphi} \right) \right]
\end{split}
\ee
We evaluate this integral analytically (see Supplementary) and find a phase-independent contribution to the current from the continuum of states:
\begin{eqnarray} \label{eq:I_cont}
I_{\text{cont}} = \frac{2 e  q v_F}{\pi \hbar}. 
\end{eqnarray} 
This part of the Josephson current is 
exactly equal to the  supercurrent that would flow in an infinite single-mode superconducting wire in the presence of finite Cooper pair momentum $q$ (a detailed discussion can be found in the Supplementary).  

\begin{figure}[t] 
	\includegraphics[width= 0.85\columnwidth]{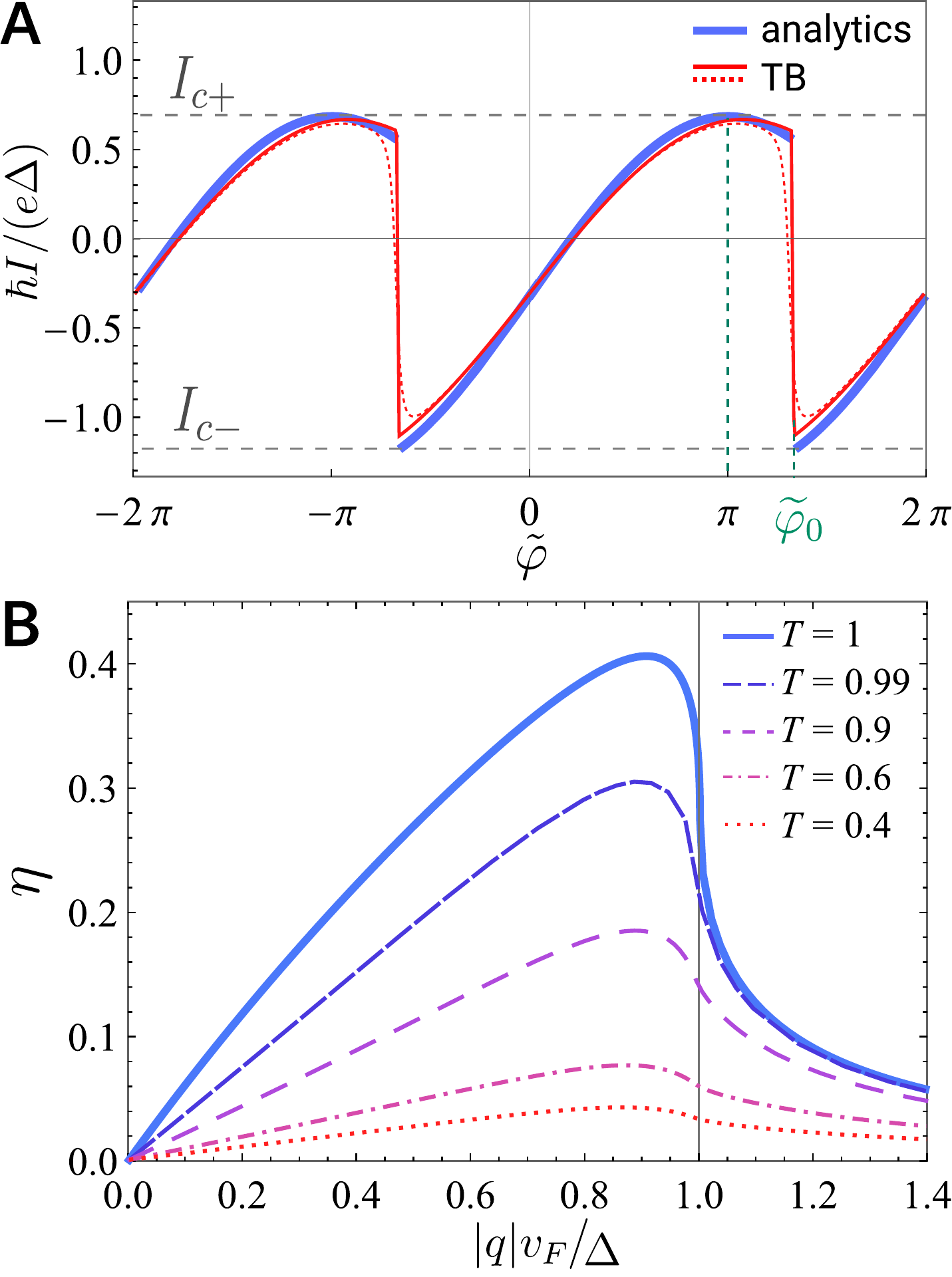}
	\caption{(\textbf{A}) Current-phase relation  at $q v_F = - 0.5 \Delta$.  The blue line is the analytical expression eq.~\eqref{J_total}; the red solid and dashed lines correspond to the tight-binding calculation for transparent junction ($T \approx 1$) and with small normal reflection ($T = 0.99$), respectively. The magnitudes of the critical currents in forward and reverse directions  are $I_{c+}$ and $I_{c-}$, respectively. 
	(\textbf{B})  The Josephson diode efficiency $\eta = |\Delta I_{c}|/(I_{c+}+I_{c-})$ as a function of the Cooper pair momentum $q$ for transparent junction ($T = 1$) and several finite values of junction transparency. 
	} \label{fig:current}
\end{figure}

Finally, the resulting total current through the junction equals
\be \label{J_total}
I =  \frac{ e \Delta}{\hbar}  \sin \frac{\tilde \varphi}{2} \text{sgn} \left( \Delta \cos \frac{\tilde \varphi}{2}   - q v_F \right) + \frac{2e  q v_F}{\pi \hbar} .
\ee
It is plotted  in Fig.~\ref{fig:current}A at $q v_F = \ -0.5 \Delta$, shown by the blue line. Because the time-reversal symmetry is broken, the antisymmetry of the current-phase relation $I(\varphi) \neq -I(-\varphi)$  doesn't hold  anymore. Additionally, when the phase bias is set to be $\widetilde \varphi \approx \varphi = 0$, there is still current flowing through the junction. This is a manifestation of the so-called `anomalous' Josephson effect~\cite{buzdin2008direct,yokoyama2013josephson,yokoyama2014anomalous,assouline2019spin,mayer2020gate,strambini2020josephson} occurring in this system.  
%This effect is a hallmark of superconducting spintronics,   and can be used to build the Josephson phase battery~\cite{strambini2020josephson}. 

Because of the identity $I =  \frac{2 e}{\hbar} \frac{d F}{d \varphi}$, the total current through the junction has to satisfy the condition $\int_0^{2 \pi} I(\varphi) d \varphi = 0$. Because the current from the bound states is modified in the presence of the Doppler energy shift, now $\int_0^{2 \pi} I_{\text{bound}}(\varphi) d \varphi \neq 0$. Interestingly, the contribution from the continuum is exactly what makes the condition of the total integral of the current being  zero satisfied. Note that in the thermodynamic equilibrium, the free energy of our system is minimized at zero current $I |_{eq}= 0$, which is attained at nonzero phase $\varphi$.

\textit{Josephson diode effect.---}
The diode effect is quantified by the difference between the magnitudes of the critical currents in the opposite directions $\Delta I_c = I_{c+} -  I_{c-}$, which equals:
\be \label{DeltaIc}
\Delta I_c = \text{sgn}(q) \left (  \frac{4 e |q| v_F}{\pi \hbar} -  \frac{e \Delta}{\hbar}\left [1  - \sqrt{1 - \left(\frac{q v_F}{\Delta} \right)^2} \right ]  \right ).
\ee
The first term is the nonreciprocal contribution from the continuum of states and the second term comes from the bound states. The contribution from the continuum of states is phase-independent, nonreciprocal, and its sign is determined by the direction of the Cooper pair momentum. The nonreciprocity from the current carried by the bound states at $E<0$ can be understood as follows: the maximum current in the positive direction %from the bound states is the largest current that  can be supported by the Andreev bound state in the junction. This 
occurs at $\widetilde \varphi = \pi$ (as in the case of $q=0$ Josephson junction), %which corresponds to the position where 
when the bound state is an equal weight superposition of right-moving electrons and left-moving holes. %its energy coincides with the middle of the Doppler shifted spectral gap for the right-movers %(the state is an equally weighted mixture of an electron and a hole at this energy).  
On the other hand, its partner formed from left-moving electrons and right-moving holes, which would carry the same current in the opposite direction, is inaccessible because it is shifted to $E>0$ by the Doppler effect. Instead, the largest current in the reverse direction occurs at the branch change $\widetilde \varphi_0$. The phases corresponding to these points are marked in Figs.~\ref{fig:spectrum}B  and \ref{fig:current}A.

The Josephson diode efficiency defined as $\eta \equiv |\Delta I_{c}|/(I_{c+}+I_{c-})$ is shown in Fig.~\ref{fig:current}A. $\eta=1$ corresponds to a perfect supercurrent diode where the critical current is 
finite in the forward direction and zero in the reverse direction. 
The maximum diode efficiency that we find reaches  about $40 \%$. This corresponds to an asymmetry of $I_{c+}/I_{c-} \approx 230$ \% which is of the order of the largest diode effects reported so far. The maximum efficiency is achieved at a universal point $q_0 v_F = \frac{4\pi}{4 + \pi^2} \Delta \approx 0.9 \Delta$. Assuming typical values of parameters $\Delta = 0.5$~meV and $\lambda_L = 140$ nm (which are close to those of bulk NbSe$_2$\cite{PhysRevLett.98.057003}) and $v_F = 10^5$ m/s, we estimate that the magnetic field $B_y \approx 40 $ mT is needed to induce optimal Cooper pair momentum $q \sim \Delta/v_F$. This is consistent with the fields used in recent observation of finite Cooper pair momentum and gapless superconductivity in ref.~\cite{zhu2021discovery}.

At small $q$, the diode effect is dominated by the  contribution from the continuum of states. Because usually the continuum contribution through the short Josephson junction vanishes, this is a remarkable example where this contribution arises due to time-reversal and inversion symmetry breaking and plays a key role.   Note that the first term in eq.~\eqref{DeltaIc} is independent of $\Delta$ and the second term only becomes large only when $q$ approaches $\Delta/v_F$. Therefore, the asymmetry between the critical currents $\Delta I_c$ provides a measure of the Doppler shift in energy, while the critical current itself is the measure of the gap $\Delta$. If one includes the reduction in the gap due to pair-breaking orbital effects or Zeeman field into consideration, the dominating continuum contribution to the diode efficiency will not change. At the same time, the critical current will decrease, which will lead to an increase in  the diode efficiency. 

At finite junction transparency $T$, the condition determining the spectrum of the states in the junction is generalized to:
\begin{equation} \label{main:T}
\begin{split}
&T \left ((r_A^+)^2 - e^{2 i q d + i \varphi} \right) \left ((r_A^-)^2 - e^{-2 i q d - i \varphi} \right) + \\ &+ (1-T) (1- r_A^- r_A^+)^2 =0.
\end{split}
\end{equation}
Using this condition and eq.~\eqref{dos}, which can be used to describe bound states as well as continuum, we compute the current-phase relations and the diode efficiency at finite junction transparency (see Supplementary section VII.A). The results are displayed in Fig.~\ref{fig:current}B and demonstrate that the effect, even though reduced, is robust and persists in the presence of normal reflection.

\begin{figure}[t] 
	\includegraphics[width= 0.85\columnwidth]{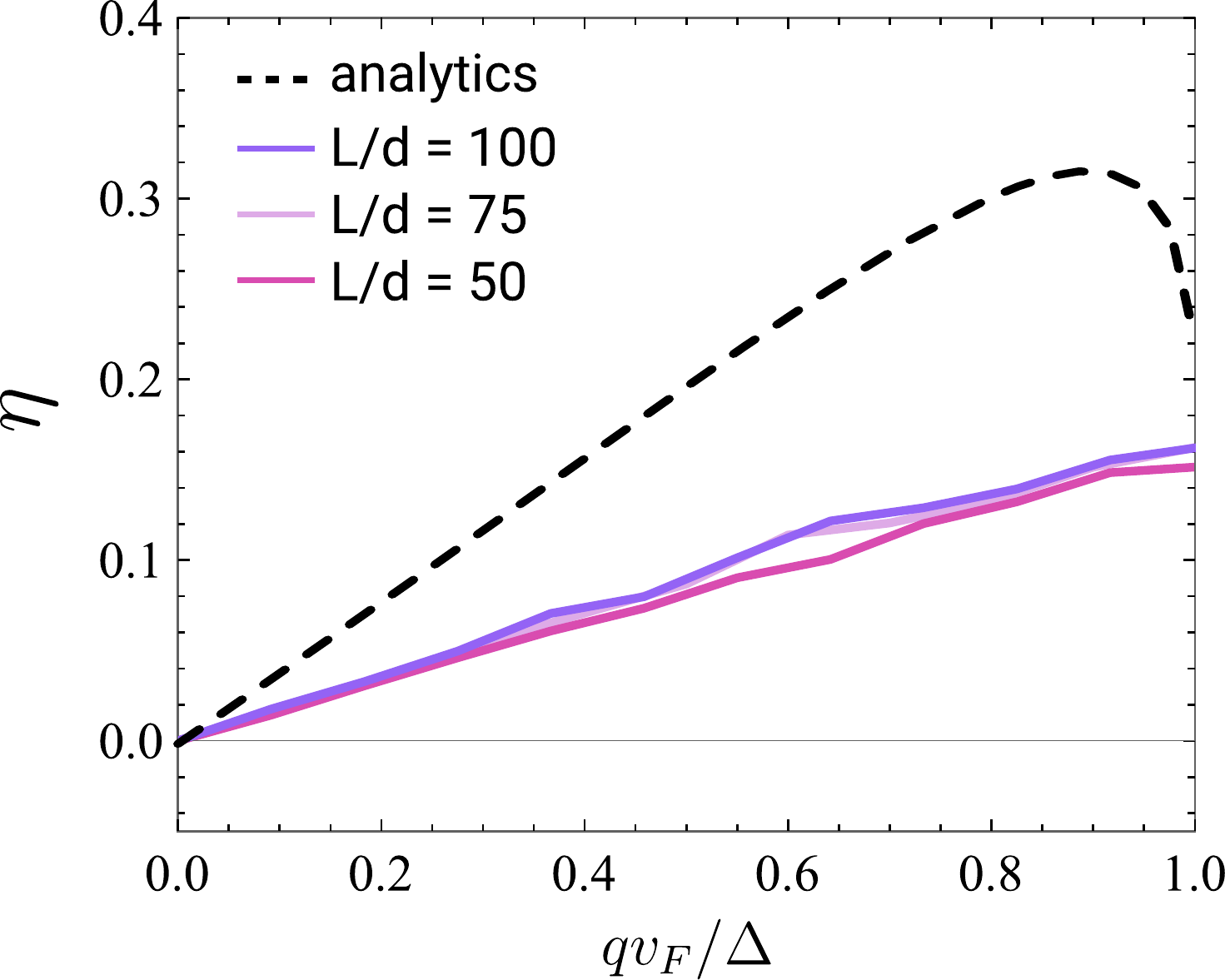}
	\caption{ The results of the tight-binding simulation of the Josephson diode effect at different  Cooper pair momenta $q$ (solid lines) averaged over disorder for different legnth of the leads $L$. The calculations were performed for the system with junction length $d = 4 a$ (where $a$ is the lattice spacing) and the length of each lead $L$. The on-site disorder in chemical potential was uniformly distributed in the range $[-10 \Delta, 10 \Delta]$. The diode efficiency was computed for each disorder realization separately and then averaged over 150 realizations. The black dashed line shows the analytical results for a clean system with $T = 0.99$ for comparison. The rest of the parameters is given in the SM. 
	} \label{fig:disorder_main}
\end{figure}

As recent experiments show~\cite{hart2017controlled,pal2021josephson}, it is easy to achieve relatively large values of Cooper pair momentum $q v_F \sim \Delta$. The situation when $|q| v_F > \Delta$ is also possible: it corresponds to gapless superconductivity in the proximitized region~\cite{yuan2018zeeman}. The gapless superconductivity is possible because the pairing potential in the nanowire is proximity-induced and thus, does not have to obey the self-consistency equation. %However, %we are not interested in this situation here, because 
In this case, the presence of mobile quasiparticles at zero energy may %lead to a finite voltage in the junction and thus 
complicate the observation of coherent phenomena, such as Josephson effect, therefore do not address this regime here (the discussion can be found in the Supplementary). For completeness, we show this domain in Fig.~\ref{fig:current}B as well.   
%In the Supplementary (Fig.~\ref{fig:currentSM}), we obtain and plot the nonreciprocal part of the current at any $q$ and show that it quickly decays to zero at $|q| v_F > \Delta$. 

The mechnism for the JDE in short junctions considered in this work is universal because it does not depend on parameters of materials in the junction and its geometry and only relies on finite Coope pair momentum. %Therefore, it can be used as a diagnostic tool for understanding the nature of the superconductivity in a system. 
This is not the case for long junctions; there, the transport occurring in the normal region becomes important; in particular, a finite-momentum Cooper pair will acquire an additional phase $\delta = 2 q d$ as it propagates through the junction. This phase shift is not small in a long junction~\cite{pal2021josephson}, and, additionally, oscillations of the diode efficiency were shown to arise due to the finite junction length. Most of the experiments so far concern the long junction limit, but the miscroscopic theory of the JDE in long junctions is a matter of future work.

\textit{The effect of normal reflection.---}
In the presence of any amount of normal reflection, there are no true bound states in the range of energies $\Delta - |q|v_F <| E |< \Delta + |q|v_F$ anymore. However, at small normal reflection, because of multiple Andreev reflections these states are quasi-bound and, when normal reflection is small, the analytical result above is still relatively accurate. We compare our analytical results for the case of transparent barrier with tight-binding calculation in Fig.~\ref{fig:current}A without and with a small barrier at the junction location (see Supplementary for the details of the calculation). %We also notice that in the case of transparent junction, the tight-binding spectrum shows that in a finite system there is only one level in the range $|E|<\Delta + |q|v_F$, which exhibits spectral flow and wraps around the domain of $\widetilde \varphi \in [0,2 \pi)$ multiple times (see Supplementary).

\textit{The effect of disorder.}--- We examine the diode effect in the presence of the chemical potential disorder in the system. We performed tight-binding simulations and have found that the diode effect is robust and persists even in the presence of relatively strong disorder. Fig.~\ref{fig:disorder_main} shows the results of the simulations of the Josephson diode in the presence of disorder uniformly distributed in the range $[-10 \Delta, 10 \Delta]$ for different lengths of the leads. In the calculations, we  assumed $\Delta = t/60$, where $t$ is the value of the next-nearest neighbor hopping and performed averaging over 150 realizations of disorder for each system length.  As we see from Fig.~\ref{fig:disorder_main} , even though the effect is reduced in magnitude, it is robust and does not depend on the system length. We find that the JDE is robust regardless of the ratio between the localization length  and the coherence length as long as $d, k_F^{-1} \ll \xi_{loc}$. Moreover, we find that when $\xi_{loc}<L$, the sensitivity of the effect to the system length disappears entirely and its magnitude depends on the values of disorder and  the superconducting gap only. 
See Supplementary Materials for more details and further analysis.

\begin{figure}[!t] 
	\includegraphics[width= 0.9\columnwidth]{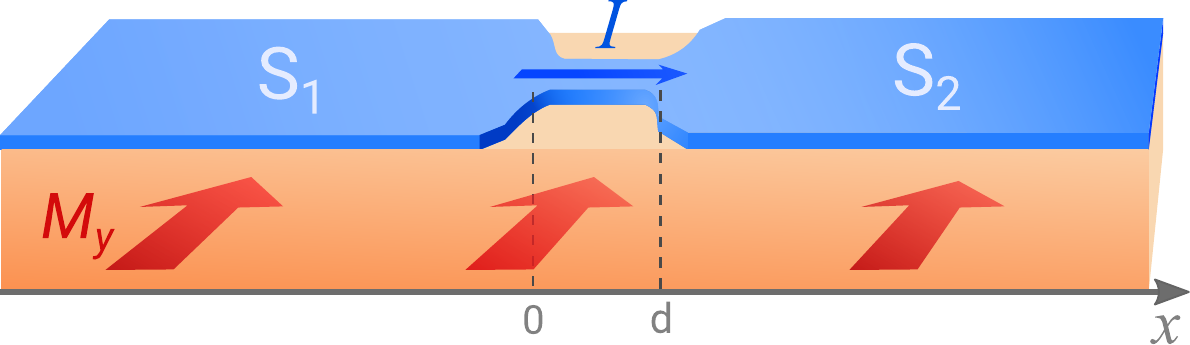}
	\caption{ Schematic illustration of  the proposal for achieving the Josephson diode effect in the absence of external magnetic field. A superconductor with a weak link at $0<x<d$ is deposited on top of a magnetic layer magnetized in $y$-direction. The finite Cooper pair momentum is achieved due to the magnetic proximity effect.  
	} \label{fig:scheme_2}
\end{figure}

\textit{Alternative ways to realize finite Cooper pair momentum.---}
Lastly, let us discuss other possible ways to realize  finite Cooper pair momentum. In Fig.~\ref{fig:scheme_2} we illustrate a superconductor-ferromagnet bilayer setup, where the magnetic proximity effect from the ferromagnet layer can be used to induce finite momentum pairing \cite{yuan2021topological} and thus achieve  the Josephson diode effect in the absence of external magnetic field. A short junction setup will realize the universal JDE mechanism proposed in this paper, which serves as a clear diagnostic tool for intrinsic finite-momentum pairing. Next, if proximitized system is a topological insulator or a semiconductor with strong spin-orbit coupling,  applying an in-plane magnetic field can lead to finite Cooper pair momentum via the Zeeman effect on spin-helical electrons  \cite{hart2017controlled,yuan2018zeeman,pal2021josephson}. Thus, constrictions based on these materials should also be a suitable platform for observing short-junction JDE. After completion of this work, we became aware of a study ~\cite{dolcini2015topological} of $\phi_0$-Josephson effect in helical edge states of a quantum spin Hall insulator, where a result  analogous to eq.~\eqref{J_total} has been found and the presence of critical current asymmetry was recognized. Note that non-magnetic impurities in the leads and potential barrier in the junction will not induce backscattering in helical edge states of the quantum spin Hall insulator, in contrast to the normal metal or semiconductor wire considered in this work.

\textit{Conclusions.---}
In this work, we developed a theory for a universal microscopic mechanism for Josephson diode effect in short Josephson junctions originating from finite Cooper pair momentum $q$. We found a large asymmetry that reaches 40 \% when $q v_F \sim \Delta$, which does not depend on the details of the junction. We also proposed a simple way to realize finite-momentum pairing based on the Meissner effect, which makes this mechanism universally applicable.

\textbf{Acknowledgements.}  We thank Noah Yuan, Banabir Pal and Stuart Parkin for related collaborations, and Anton Ahkmerov for helpful comments on the manuscript. We are grateful for useful discussions with Yasen Hou and Jagadeesh Moodera. This work is partly supported by the David and Lucile Packard Foundation.

\bibliography{ref}

\clearpage
\pagebreak
\onecolumngrid
\begin{center}
%\textbf{\large Supplemental Materials}
\end{center}

\setcounter{equation}{0}
\setcounter{figure}{0}
\setcounter{table}{0}
\setcounter{page}{1}
\makeatletter
\renewcommand{\theequation}{S\arabic{equation}}
\renewcommand{\thefigure}{S\arabic{figure}}

\vspace{0 pt}

\section{Scattering states}

We linearize the problem and consider $+/-$  (where $+/-$ corresponds to the vicinity of $\pm k_F$, i.e. right- and left-movers) and use the ansatz
\be
\psi_{1,2} = \begin{pmatrix}
a^+_e e^{i k_F x} e^{i(k+q)x + i \varphi_{1,2}/2}\\
a^+_h e^{i k_F x} e^{i(k-q)x - i \varphi_{1,2}/2}\\ a^-_e e^{-i k_F x} e^{i(k+q)x + i \varphi_{1,2}/2}\\ 
a^-_h e^{-i k_F x} e^{i(k-q)x - i \varphi_{1,2}/2}
\end{pmatrix}
\ee
in superconducting lead 1 and similarly in lead 2. Here $k \equiv k_x$.  Recall that $\varphi_1 = 0$ and $\varphi_2 = \varphi$.
Notice that normally, there should be wavefunction normalization by the quasiparticle current in order for the scattering problem to be unitary; however, we will use Andreev approximation, which makes this normalization unnecessary.

We assume that $0< q v_F < \Delta$, and that the energies of the states will lie in the gap of the superconductor $0<E< \Delta - q v_F$ .

The wavefunction for right-movers that decays as $x\rightarrow + \infty$ in $SC_2$:
\be
 \begin{pmatrix}
a^+_e \\
a^+_h \\ 
0\\ 
0
\end{pmatrix}_{SC_2} = \frac{1}{\sqrt{2}} \begin{pmatrix}
\left( \frac{E- v_F q}{\Delta} + i \sqrt{1-\left( \frac{E- v_F q}{\Delta}\right)^2}\right)\\
1\\ 
0\\ 
0
\end{pmatrix} 
\ee
For this state, $v_F k =+ i \sqrt{\Delta^2 - (E- v_F q)^2}$.

The L state in $SC_1$ that decays as $x\rightarrow -\infty$ is:
\be
 \begin{pmatrix}
0 \\
0 \\ 
a^-_e\\ 
a^-_h
\end{pmatrix}_{SC_1} =
\frac{1}{\sqrt{2}} 
\begin{pmatrix}
0\\
0\\
\left( \frac{E+ v_F q}{\Delta} + i \sqrt{1-\left( \frac{E+ v_F q}{\Delta}\right)^2}\right)\\
1
\end{pmatrix} 
\ee
For this state, $v_F k =- i \sqrt{\Delta^2 - (E+ v_F q)^2}$.

It is known~\cite{beenakker1992three} that the result of the  scattering formalism for short junctions will be independent of whether the junction is represented by narrow weak link, a region of normal metal, or an insulating barrier. For simplicity of calculation, we consider normal states in the middle region. 
We will be solving the problem of Andreev scattering at two interfaces ($N/SC_2$ and $SC_1/N$). For incoming electron, the scattering states with that are relevant to  the left (1) and the right (2) contacts are:
\be
\psi_N^{(1)} = \begin{pmatrix}
0 \\
0\\ 
e^{-i k_F x} e^{i k x}\\ 
r_A  e^{-i k_F x} e^{i k x}
\end{pmatrix}, \quad \psi_N ^{(2)} = \begin{pmatrix}
 e^{i k_F x}  e^{i k x} \\
r_A e^{i k_F x} e^{i k x} \\ 
0 \\ 
0 \end{pmatrix}
\ee
Where $r_A$ is not necessarily the same constant for scattering on the left and right contacts. For the incoming holes, the problem is set up similarly.

\newpage
\section{The scattering matrix formalism}

To obtain the amplitudes of Andreev reflection, we solve the condition $\psi_N^{(1,2)} = S_{interface \ 1,2} \psi_S^{(1,2)}$ at each interface ($x = 0$ and $x=d$). In the case of perfectly transparent contacts, the scattering matrix at the interfaces is identity. We solve these equations (for both incoming electrons and holes at both interfaces) and obtain the matrix describing Andreev scattering at both interfaces:
\be \begin{split}
\psi_{out} = \begin{pmatrix}
\psi_{N,e}^-(0)\\ 
\psi_{N,e}^+(d)\\ 
\psi_{N,h}^+(0)\\ 
\psi_{N,h}^-(d)
\end{pmatrix}  = \begin{pmatrix}
 &  & r_A^- & 0\\ 
 &  & 0 & r_A^+ e^{-  i \varphi}\\ 
r_A^+ & 0 &  &  \\
0 & r_A^- e^{  i \varphi}  &   &  
\end{pmatrix}  \begin{pmatrix}
\psi_{N,e}^+(0)\\ 
\psi_{N,e}^-(d)\\ 
\psi_{N,h}^-(0)\\ 
\psi_{N,h}^+(d)
\end{pmatrix}\equiv s_A^{-1} \psi_{in}
\end{split}
\ee
where the unfilled spaces correspond to zero entries,  and we used the notation
\begin{align}
r_A^{\pm} = \frac{E \mp v_F q}{\Delta } - i \sqrt{1 - \left (\frac{E\mp v_Fq}{\Delta} \right)^2}.
\end{align}
The scattering matrix of the normal region is:
\be
\psi_{out} =\begin{pmatrix}
\psi_{N,e}^-(0)\\ 
\psi_{N,e}^+(d)\\ 
\psi_{N,h}^+(0)\\ 
\psi_{N,h}^-(d)
\end{pmatrix}   = \begin{pmatrix}
 r& t' &  & \\ 
t & -r' &  & \\ 
 &  & r^* & t'^* \\
 &  &  t^* & -r'^* 
\end{pmatrix} \begin{pmatrix}
\psi_{N,e}^+(0)\\ 
\psi_{N,e}^-(d)\\ 
\psi_{N,h}^-(0)\\ 
\psi_{N,h}^+(d)
\end{pmatrix}  \equiv s_N \psi_{in}
\ee
where in the limit of short junction ($\frac{\Delta d}{v_F} \approx \frac{d}{\xi}, \frac{E_z d}{v_F} \ll 1$) the transmission and reflection are energy-independent.

In our case, for one channel, $r' = r$ and $t' = t$. The condition determining the spectrum of the ground states is $\det \left ( \bm 1 - s_N s_A \right) = 0 $,  which translates into
\be \label{T}
T \left ((r_A^+)^2 - e^{2 i q d + i \varphi} \right) \left ((r_A^-)^2 - e^{-2 i q d - i \varphi} \right) + (1-T) \left [(1- r_A^- r_A^+)^2  \right ]=0.
\ee
where $T = |t|^2$, $|t|^2 + |r|^2 = 1$. In the absence of normal reflection $t = e^{i qd}$, $T = 1$ and this simplifies to:
\be \label{main}
\left ((r_A^+)^2 - e^{2 i q d + i \varphi} \right) \left ((r_A^-)^2 - e^{-2 i q d - i \varphi} \right) =0
\ee
from the main text, and the solutions to this equation produce the energies of the two bound states \eqref{eq:ABS_energy}.
%%%%%%%%%%%%%%%%%%%%

\begin{figure}[h] 
	\includegraphics[width= 0.9\columnwidth]{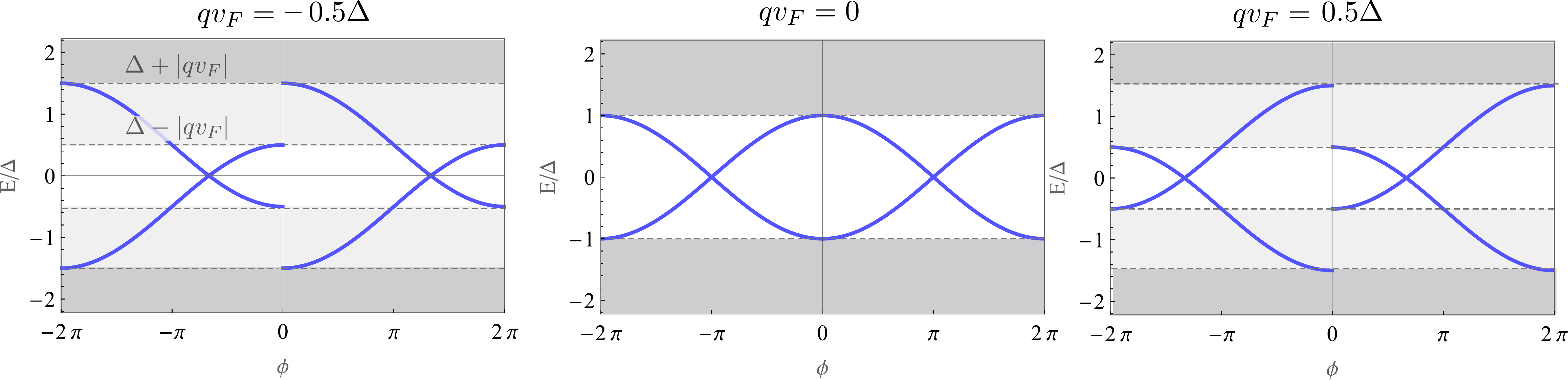}
	\caption{ Spectrum of the bound states in the junction at different values of the Cooper pair momentum $q$ in superconducting regions 1 and 2. 
	} \label{Espectrum}
\end{figure}

\section{The current-phase relation}

The free energy is \cite{beenakker1991universal,beenakker1992three}
\be
F = -\frac{2}{\beta} \sum_{E>0} \ln \left [ 2 \cosh \left( \frac{\beta E}{2} \right )\right] + \int d^2 r \frac{|\Delta|^2}{|g|} + \Tr H_0
\ee

where $H_0$ is the particle block of the BdG Hamiltonian and $\beta$ is the inverse temperature. We neglect the contribution from  the spatial integral $ \int d^2 r \frac{|\Delta|^2}{|g|}$ to the Josephson current because we assume that $\Delta$ changes as a step-function at the contacts.  Therefore, the free energy can be written as
\be
F = - \frac{1}{\beta} \int_0^\infty dE \nu(E)  \ln 2\cosh\frac{\beta E}{2}
\ee

The density of states $\nu(E)$ in the junction  can be evaluated as\cite{beenakker1992three}, 
\be \label{Supp:nu}
\nu(E)=-\frac{1}{\pi } \Im \frac{\partial}{\partial E} \ln \det\left ( 1- s_A s_N\right)  + \text{const.}
\ee
which is good for describing both bound and continnum states. Here $s_A$ and $s_N$ are the scattering matrices transforming the wavefunctions due to Andreev reflection at the interfaces and due to the propagation/scattering in the weak link; the `const.' is the phase-independent part of the density of states.

In the absence of normal reflection, we can rewrite the density of states as
\be
\nu(E)=\frac{1}{\pi } \Im \frac{\partial}{\partial E} \ln \sin\left (   \arccos\frac{E + q v_F}{\Delta} + \frac{\widetilde \varphi}{2}\right)\sin\left (   \arccos\frac{E - q v_F}{\Delta} - \frac{\widetilde \varphi}{2}\right)  + \text{const.}
\ee
where, in order to work with energies of both bound and continuous states, we have to assume that $E = E + i 0$, and perform proper analytic continuation where necessary.

Lastly, we plug the expression for the density of states into the free energy and evaluate the current as $I =  \frac{2 e}{\hbar} \frac{d F}{d \varphi}$. We extend the symmetric integration to $(-\infty, +\infty)$ and integrate by parts using that the boundary terms vanishing as $\propto 1/E$. Thus, we obtain:
\be
I(\varphi) = - \frac{e}{2 \pi \hbar} \int_{-\infty}^{\infty} dE \tanh \frac{\beta E}{2} \mathrm{Im} \frac{\partial}{\partial \varphi} \ln  \sin\left (   \arccos\frac{E + q v_F}{\Delta} + \frac{\widetilde \varphi}{2}\right)\sin\left (   \arccos\frac{E - q v_F}{\Delta} - \frac{\widetilde \varphi}{2}\right) 
\ee
which is equal 
\be
I(\varphi) = - \frac{e}{4 \pi \hbar} \int_{-\infty}^{\infty} dE \tanh \frac{\beta E}{2} \mathrm{Im} \left [ \cot\left (   \arccos\frac{E + q v_F}{\Delta} + \frac{\widetilde \varphi}{2}\right) - \cot\left (   \arccos\frac{E - q v_F}{\Delta} - \frac{\widetilde \varphi}{2}\right)  \right ]
\ee
We complete the contour in the upper half complex energy plane, picking up residues at each of the poles of the hyperbolic tangent at  Matsubara frequencies $E = i\omega_n  \equiv i(2n+1)\pi/\beta$ to obtain a summation:
\be \label{supp_I}
I(\varphi) = \frac{e}{\hbar \beta} \mathrm{Re} \left [ \cot\left (   \arccos\frac{i \omega_n + q v_F}{\Delta} + \frac{\widetilde \varphi}{2}\right) - \cot\left (   \arccos\frac{i \omega_n  - q v_F}{\Delta} - \frac{\widetilde \varphi}{2}\right)  \right ]
\ee
Which, finally, can be brought into the form:

 \begin{equation}
	I(\varphi) = \frac{2e}{\beta \hbar} \sum_{n=0}^{\infty}  \textrm{Re}\cot \left(  \arccos(\frac{i\omega_n-qv_F}{\Delta}) - \frac{\tilde \varphi}{2}\right)
\label{eqn:sum}
\end{equation}

The zero temperature result, which we are interested in, is given by turning the sum $\sum_{n=0}^{\infty}$ into an integral $\frac{\beta}{2\pi}\int_0^\infty d\omega $.

\subsubsection{ The case of $|q| v_F < \Delta$}

\begin{figure}[h] 
	\includegraphics[width= 0.4\columnwidth]{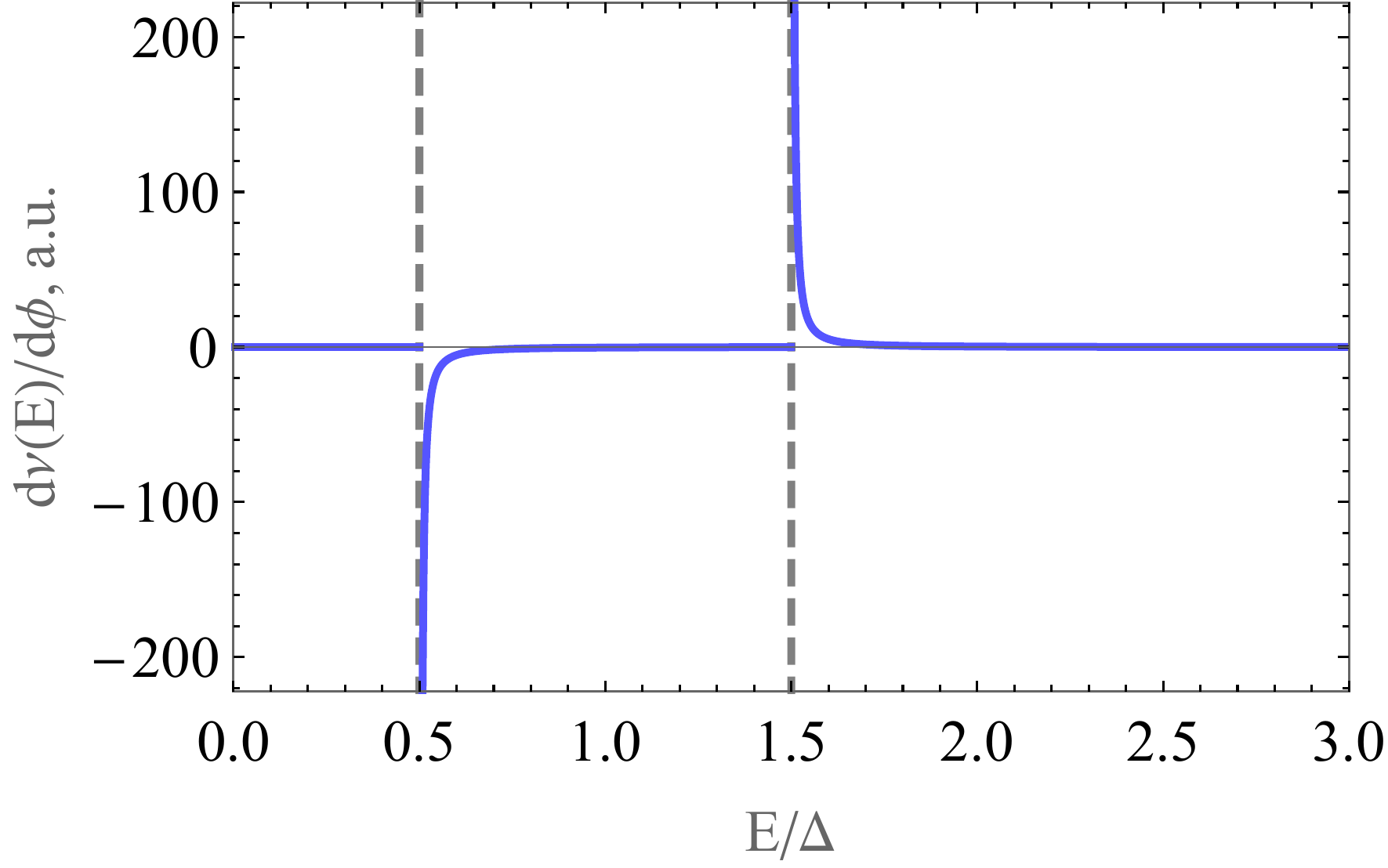}
	\caption{  The derivative over $\varphi$  of the density of states for the continuum part of the spectrum only at $q v_F = 0.5 \Delta$. The dashed gray lines correspond to $E = \Delta - q v_F$ and $E = \Delta + q v_F$. The tails of the quantity $d \nu (E)/ d\varphi$ in the continuum have the opposite sign and cancel exactly when $q = 0$. When $q \neq 0$, the energies corresponding to the continuum states for right- and left-movers acquire Doppler shift $\pm q v_F$; on top of that, there is no perfect cancellation anymore, as we show in the calculation above.
	} \label{fig:dos}
\end{figure}

Here we derive the Josephson current at zero temperature using a slightly simpler approach than the one introduced in the section above. This approach will also allow us to distinguish the current contributions from bound states and continuous states. The Josephson current can be written as:
\be
\begin{split}
I = -  \frac{2e}{\hbar}\sum_{E>0} \tanh \left( \frac{\beta E}{2 } \right ) \frac{dE}{d\varphi} -  \frac{2e}{\hbar}\frac{2}{\beta} \int     _{E\,  \in \text{ cont.}}^\infty dE \ln \left [ 2 \cosh \left( \frac{\beta E}{2} \right )\right] \frac{d \nu (E)}{d \varphi} 
\end {split}
\ee
Where the first term only counts contributions from the bound states and the second term corresponds to the current from the continuum. We are interested in the case when the temperature is zero, and the expression simplifies:
\be
\begin{split}
I = -  \frac{2e}{\hbar}\sum_{E>0} \frac{dE}{d\varphi} -  \frac{2e}{\hbar} \int     _{E\,  \in \text{ cont.}}^\infty dE \, E  \frac{d \nu (E)}{d \varphi} 
\end {split}
\ee
The expression for the contribution from the bound states is given in the main text (eq. \eqref{J_bound}). In what follows, we present the details of derivation of the current arising from the continuum of states. $\nu(E)$ equals
\be
\nu(E)=-\frac{1}{\pi } \Im \frac{\partial}{\partial E} \ln  \left ( T \left ((r_A^+)^2 - e^{2 i q d + i \varphi} \right) \left ((r_A^-)^2 - e^{-2 i q d - i \varphi} \right) + (1-T) (1- r_A^- r_A^+)^2\right )
\ee
When $T = 1$, the expression is especially simple. The derivative over $\varphi$ of the density of states for the continuum of states at $q v_F = 0.5 \Delta$ and $\varphi = 0$ is shown in fig.~\ref{fig:dos}.

To compute the current from the continuum of states analytically, we change the order of derivatives and separate the contributions from left- and right-movers:
\be
I_{cont} = \frac{2e}{\hbar} \frac{1}{\pi} \Im \left [\int_{\Delta + |q|v_F}^\infty dE \, E \frac{\partial }{\partial E} \frac{d}{d\varphi} \ln \left ( (r_A^+)^2 - e^{ i \tilde \varphi} \right)+ \int_{\Delta - |q|v_F}^\infty dE \, E \frac{\partial }{\partial E} \frac{d}{d\varphi} \ln \left ( (r_A^-)^2 - e^{- i \tilde \varphi} \right) \right]
\ee
Then we proceed to evaluate the derivative over $\varphi$:
\be
I_{cont} = -\frac{2e}{\hbar} \frac{1}{\pi} \Im i\left [\int_{\Delta + |q|v_F}^\infty dE \, E \frac{\partial }{\partial E}  \frac{1}{  e^{ -i \tilde \varphi} (r_A^+)^2 -1 }- \int_{\Delta - |q|v_F}^\infty dE \, E \frac{\partial }{\partial E}\frac{1}{  e^{ i \tilde \varphi} (r_A^-)^2 -1 } \right]
\ee
Next, we integrate  by parts in order to obtain
\be
\begin{split}
I_{cont} &= \frac{2e}{\hbar} \frac{1}{\pi} \Im i\left [(\Delta - |q|v_F) \frac{1}{  e^{ -i \tilde \varphi}-1} - (\Delta + |q|v_F)  \frac{1}{ e^{ -i \tilde \varphi}-1 }\right] +\\
&+\frac{2e}{\hbar} \frac{1}{\pi} \Im i\left [\int_{\Delta + |q|v_F}^\infty dE \, \frac{1}{  e^{ -i \tilde \varphi} (r_A^+)^2 -1}- \int_{\Delta - |q|v_F}^\infty dE \,\frac{1}{  e^{ i \tilde \varphi} (r_A^-)^2 -1 } \right]
\end{split}
\ee
Note that $\Im  \frac{i}{ e^{ -i \tilde \varphi}-1 }=\Im  \frac{i}{ e^{ i \tilde \varphi}-1 }=\frac{1}{2}$. Thus, simplifying further:
\be
\begin{split}
I_{cont} &=-\frac{e}{\hbar} \frac{1}{\pi} \left [(\Delta - |q|v_F)  - (\Delta + |q|v_F) \right] 
+\frac{2e\Delta }{\hbar} \frac{1}{\pi} \Im i\left [\int_{1}^\infty dx \, \frac{1}{  e^{ -i \tilde \varphi + 2 \text{arccosh} x}  -1 }- \int_{1}^\infty dy \,\frac{1}{  e^{ i \tilde \varphi+2 \text{arccosh} y} -1 } \right] = 
\\
&=\frac{e}{\hbar} \frac{2 |q| v_F}{\pi}  - \frac{2e\Delta }{\hbar} \frac{1}{\pi} \Im \left [\int_{1}^\infty dx \, \frac{2  e^{2 \text{arccosh} x} \sin \tilde \varphi }{  1+ e^{4 \text{arccosh} x}  -2 e^{2 \text{arccosh} x} \cos \tilde \varphi} \right]
\end{split}
\ee
(recall that for continuous states $(r^{\pm}_A)^2 = e^{2 \text{arccosh} \frac{E \mp q v_F}{\Delta}}$). We see that the imaginary part of the  second term is identically zero and thus:
\be
\begin{split}
I_{cont} =\frac{e \Delta}{\hbar} \frac{2 q v_F}{\pi \Delta}  
\end{split}
\ee
which yields the result in eq.~\eqref{eq:I_cont}.

\subsubsection{The case $|q|v_F > \Delta$}
When $qv_F >\Delta$ and $0<\widetilde{\varphi}<2\pi$, we evaluate the current using eq.~\eqref{eqn:sum} at zero temperature:
 \begin{equation}
	I(\varphi) = \frac{e}{\pi \hbar} \int_{0}^{\infty}  d\omega  \textrm{Re}\cot \left(  \arccos(\frac{i\omega -qv_F}{\Delta}) - \frac{\tilde \varphi}{2}\right)
\label{eqn:sum1}
\end{equation}
We obtain:
\begin{eqnarray} I(\varphi) = -\frac{2e(qv_F - \sqrt{q^2v_F^2 - \Delta^2})}{\pi}  + 
 \frac{2e\Delta \sin(\frac{\widetilde{\varphi}}{2})}{2\pi}
 ( \pi - \textrm{arg}[\Delta + qv_F\cos(\frac{\widetilde{\varphi}}{2}) 
 +i\sqrt{q^2v_F^2 - \Delta^2}\sin(\frac{\widetilde{\varphi}}{2})] \\ -
 \textrm{arg}[\Delta - qv_F\cos(\frac{\widetilde{\varphi}}{2}) + i\sqrt{q^2v_F^2 - \Delta^2}\sin(\frac{\widetilde{\varphi}}{2}) ])   \end{eqnarray}
 Which can be simplified to 
 \be I(\varphi) = -\frac{2e(qv_F - \sqrt{q^2v_F^2 - \Delta^2})}{\pi}  + 
 \frac{2e\Delta \sin(\frac{\widetilde{\varphi}}{2})}{\pi} \arctan{ \frac{\Delta \sin{\frac{\widetilde{\varphi}}{2}}}{\sqrt{q^2v_F^2-\Delta^2}}}
 \ee
 Here $\textrm{arg}(z)$ refers to the argument of the complex number $z$. %Note that this formula does not hold for $2\pi < \phi < 4\pi$; in that range $I(\phi) = I (\phi -2\pi)$
 
 The maximum negative current occurs at $\widetilde{\varphi} = 0$, $|I_{c-}| = \frac{2e}{\pi}(qv_F - \sqrt{q^2v_F^2 - \Delta^2})$.
 The maximal positive current occurs at $\widetilde{\varphi} = \pi$, $I_{c+} = \frac{2e}{\pi}(\Delta \sin^{-1}(\Delta/qv_F)-(qv_F - \sqrt{q^2v_F^2 - \Delta^2}))$. Thus, the  diode efficiency can be expressed as
 $\frac{I_{c-} - I_{c+} }{I_{c-} + I_{c+}} = \frac{  2(1-\sqrt{1-p^2})-p\sin^{-1}(p)}{p\sin^{-1}(p)}$, where $p \equiv \Delta/qv_F <1$.
 
 In the limit $qv_F \gg \Delta$, the diode efficiency  approaches $p^2/12 + 13p^4/360 + \ldots$ and thus vanishes as $\propto 1/q^2$. The expression for the supercurrent becomes symmetric in this limit $I(\varphi)|_{q v_F/\Delta \rightarrow + \infty} \approx -\frac{e\Delta^2 \cos{\widetilde{\varphi}}}{\pi qv_F} $.

\begin{figure}[h] 
	\includegraphics[width= 0.35\columnwidth]{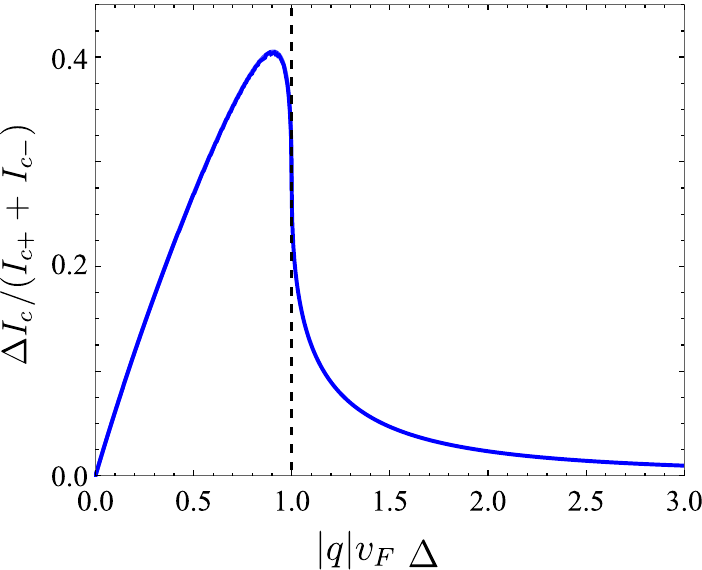}
	\caption{  The diode efficiency  $\eta$ through the junction as a function of the Cooper pair momentum $q$. 
	} \label{fig:currentSM}
\end{figure}

\subsubsection{Magnetic field at an angle to the junction}
\begin{figure}[t] 
	\includegraphics[width= 0.5\columnwidth]{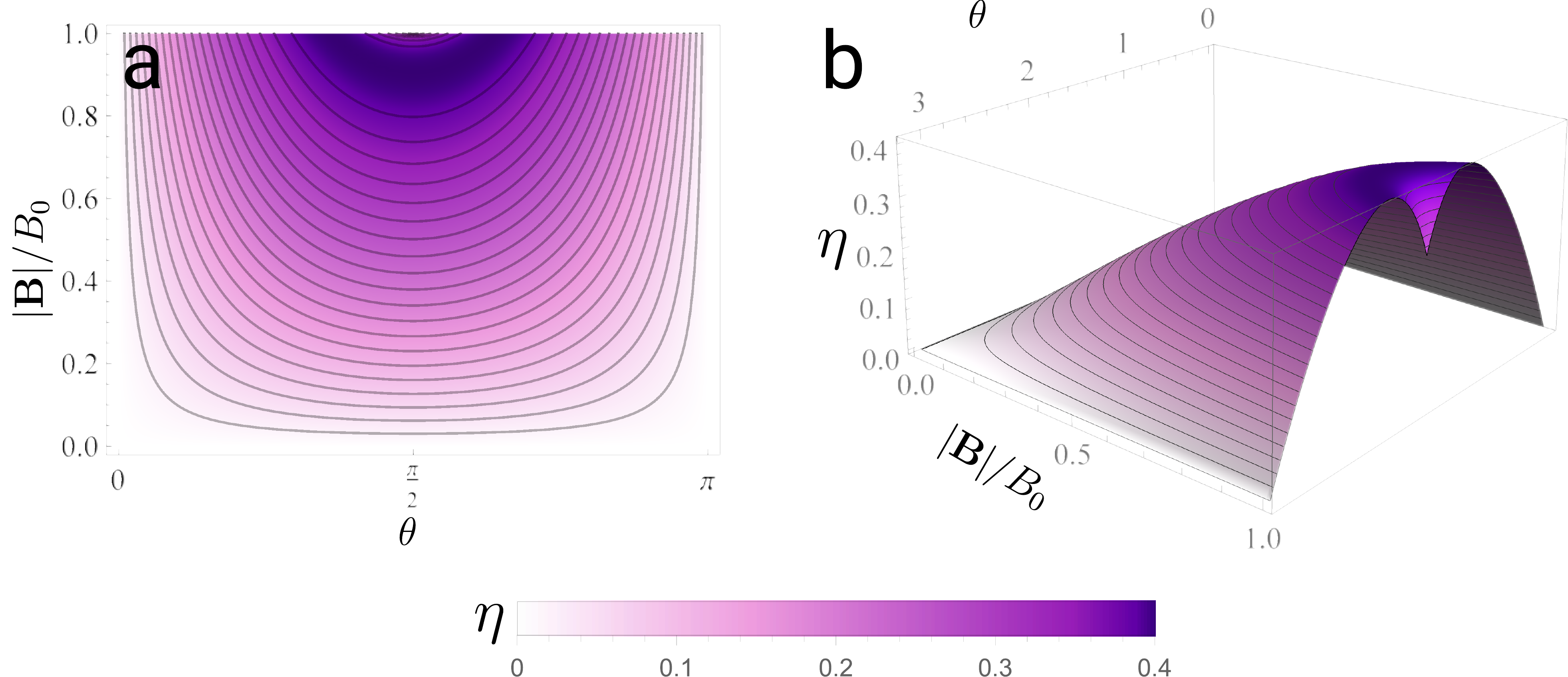}
	\caption{Dependence of the diode efficiency on the magnitude of magnetic field $|\bm B|$ and the angle  $\theta$ between the magnetic field and the direction of the current in the domain of parameters satisfying condition $q_x < \Delta/v_F$. $B_0$ is the value of magnetic field at which  $q_x = \Delta/v_F$.
	} \label{fig:SM_Btheta}
\end{figure}

In a quasi-1D geometry when there are only few transverse modes in the junction, the effect of misaligning the in-plane magnetic field $\bm B$ from the direction $y$ perpendicular to the current is easy to understand. The effect of the $x$-component of the field, which is in-plane and parallel to the junction leads only to Zeeman energy. At small magnetic fields, this contribution can be neglected, and therefore the Cooper pair momentum in $x$-direction equals $|\bm B| \sin \theta$, where $\theta$ is the angle between the current and direction of the magnetic field. The plot of this dependence is shown in Fig.~\ref{fig:SM_Btheta}.

\section{Tight-binding calculations}

We simulate the minimal model with short Josephson junction by setting up a nearest-neighbor tight-binding chain with  superconducting regions of the same length $L_S = N_S a$ and the thickness of the normal region $L_N = N_N a$, $N_N \ll N_S$. The hopping amplitude $t$ is the same in all regions, and the chemical potentials are the same in the superconducting regions $\mu_S$ and $\mu_N$ in the normal region. The pairing potential at lattice site $n$ is $\Delta(n) = \Delta_{1,2} e^{2 i q n a}$, where $\Delta _1 = \Delta$ and $\Delta_2 = \Delta e^{i \varphi}$. When $t \gg \Delta$, this corresponds to the condition $\mu \gg \Delta$ used in analytical derivations. In all the calculations, we used $\mu_S = 0$, and thus, $v_F = 2 a t$.

\begin{figure}[h] 
	\includegraphics[width= 1\columnwidth]{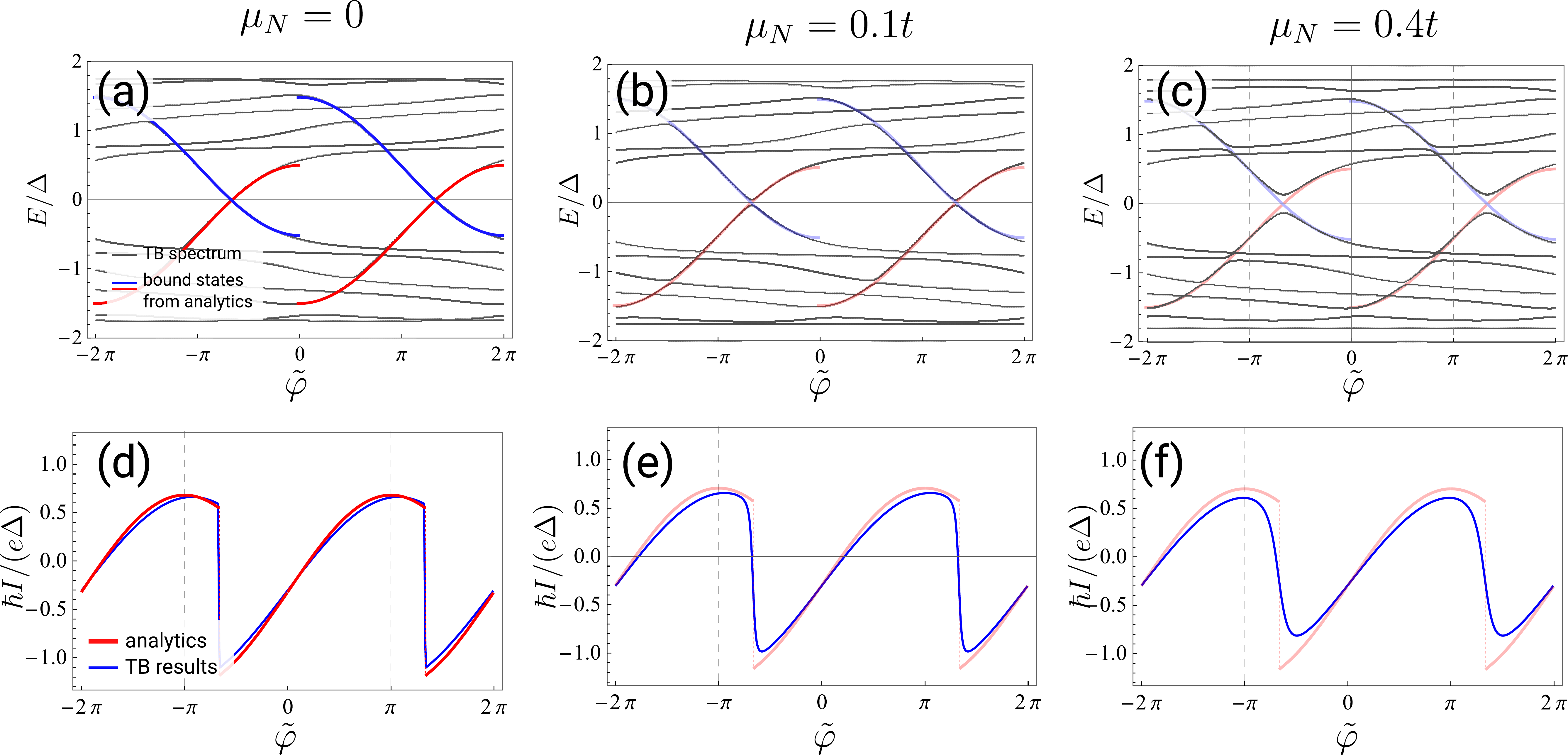}
	\caption{(a-c) Energy spectra of bound and extended states from tight-binding calculation at $q v_F = -0.5 \Delta$. The parameter $\mu_N$ is the potential at the junction that controls the amplitude of the normal reflection.  The other parameters used are $N_S = 150$, $N_N = 3$, $a = 1$, $t = 20$, $\Delta = 2$. (d-f) Corresponding phase-current relations showing that the current nonreciprocity is decreased when the normal reflection becomes large.
	} \label{fig:SM_TB_spectrum}
\end{figure}

For calculation in Fig.~\ref{fig:current}, we used $N_S = 350$, $N_N = 3$ (the total length of the system is 703$a$), $a = 1$, $t = 100$ and $\Delta = 2$. The solid line shows the result at negligible normal reflection, which is achieved at $\mu_N = 0$. The dotted line shows the result at small normal reflection, when a small potential barrier is introduced inside the junction by setting $\mu_N = 25$, which opens a small gap in the dispersion, see Fig.~\ref{fig:SM_TB_spectrum}(a-c). 

The current was found by  evaluating the expression $I =  \frac{2 e}{\hbar} \frac{d F}{d \varphi}$ numerically, where the free energy is found by summing over all the negative energy states. We plot it in Fig.~Fig.~\ref{fig:SM_TB_spectrum}(d-f). We estimate that when the potentials at the junction are equal to $\mu_N = 0.1t$ and $\mu_N = 0.4 t$, the junction transparency is $T = 0.998$ and $T = 0.975$, respectively. We obtained this correspondence by comparing the energy spectrum obtained from the tight-binding calculation with the analytical expression $E = \sqrt{1- T \sin^2 \frac{\varphi}{2}}$. We estimate the junction transparency to be $T = 0.99$ for the tight-binding calculation at $\mu_N = 0.25 t$ shown as a dashed red line in Fig.~\ref{fig:current}A.

\section{Spectral flow}

\begin{figure}[h] 
	\includegraphics[width= 0.6\columnwidth]{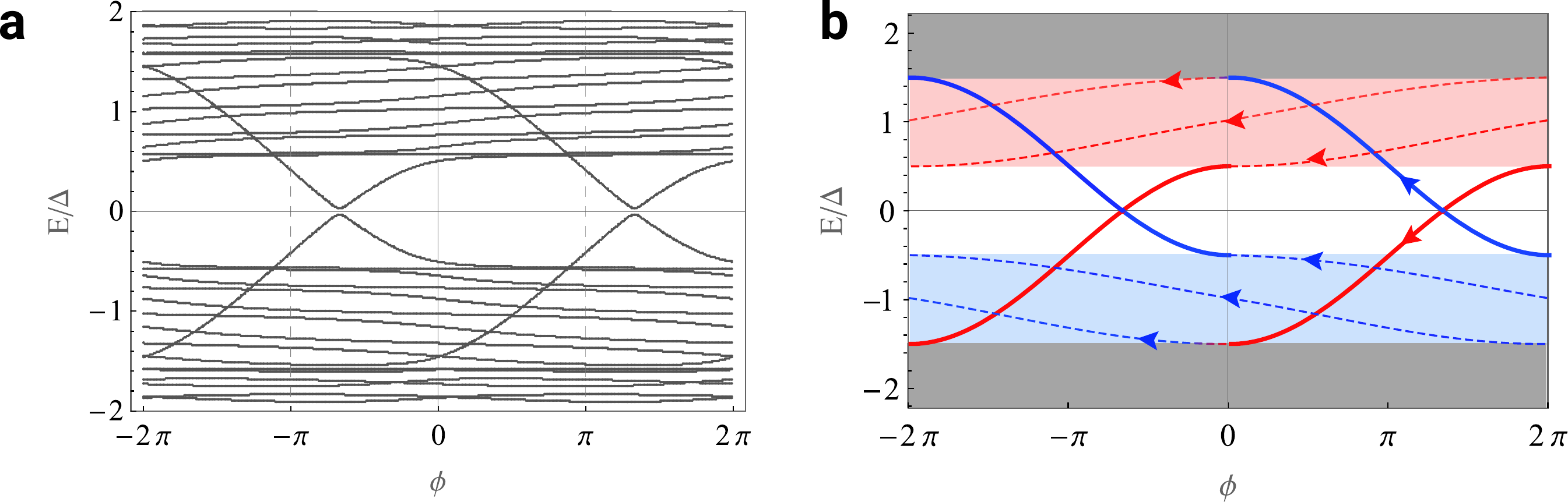}
	\caption{(a)Tight-binding calculation of the spectrum of a 1D nanowire at $q v_F = - 0.5 \Delta$. . The parameters used are $N_S = 150$, $N_N = 3$, $a = 1$, $t = 20$, $\Delta = 2$, and $\mu_N = 0$. (b) Analytical expression for bound state levels at $T = 1$ and schematic illustrating the spectral flow connecting the states.
	} \label{Espectrum_TB}
\end{figure}

In the presence of normal reflection, the left-and right moving states mix, and in the energy domain 
$\Delta - |q| v_F < |E|< \Delta + |q| v_F$ there are no true bound states anymore.  
As we see, now the contributions from left-movers and right-movers at these energies (associated with $r_A^+$ and $r_A^-$, respectively) are now related.
From tight-binding calculations, we see that these states are not connected to the rest of the continuum states as shown in Fig. \ref{Espectrum_TB}. Upon changing the phase, there is spectral flow of one bound state into these quasi-continuum states and back into another bound states. This allows us to estimate the contribution of the continuum states into the Josephson current based on spectral flow argument:
\be \label{cont2}
I_{cont}= - \frac{2 e}{\hbar} \sum_{i \text{ in continuum}} \frac{d |E_i|}{d \varphi} =  \frac{2e }{\hbar } \frac{\Delta E}{\Delta \varphi } = -\frac{2e }{\hbar } \frac{2 |q| v_F}{2 \pi} = -\frac{e \Delta }{\hbar } \frac{2 |q| v_F}{ \pi \Delta}
\ee
Which exactly matches the result in Eq. \eqref{eq:I_cont}. 

\section{Further discussion of the contribution from the continuum of states}

\subsubsection{Vanishing of the contribution from continuum in conventional short junctions}

Let us discuss one perspective on how and why continuous spectrum at $\Delta - |q v_F|<E<\Delta +|q v_F|$ contributes to the Josephson current.
As we saw above, the current from continuous states is determined from 
\be \label{Icont_gen}
I_{cont} =  -  \frac{2e}{\hbar} \int     _{\text{ cont.}} dE \, E  \frac{d \nu (E)}{d \varphi}, \quad %
\nu(E)=-\frac{1}{\pi } \Im \frac{\partial}{\partial E} \ln \det\left ( 1- s_A s_N\right)  + (\varphi \text{-independent const})
\ee
Consider
\be \label{det_DOS}
\det\left ( 1- s_A s_N\right) =   T \left ((r^-_A)^2 (r^+_A)^2 +1 - 2 \cos \tilde \varphi \left [ (r^-_A)^2 + (r^+_A)^2\right] \right) - T  (1-T) (1- r_A^- r_A^+)^2 - 2 i \sin \tilde \varphi \left [ (r^-_A)^2 - (r^+_A)^2\right] 
\ee
In the cases considered in refs.\cite{beenakker1992three,furusaki1991,beenakker1991universal}, the dispersion for left- and right-movers was symmetric and  for short junction:
\be
r^-_A = r^+_A = r_A
\ee
Which immediately sets imaginary part of the determinant above zero. Thus, for a short junction with L/R-symmetric dispersion, the density of states is independent of $\varphi$ and the current from the continuum \eqref{Icont_gen} vanishes.

It is known that just time-reversal symmetry breaking (for example, induced by spin-splitting magnetic field, see \cite{yokoyama2013josephson,yokoyama2014anomalous}) does not lead to an asymmetry in ABS dispersion in the case of short Josephson junctions. Therefore, the fact that the finite Cooper pair momentum not only breaks time-reversal, but also provides a selected direction in space (inversion breaking) is crucial for the asymmetry and the JDE effect.

\subsubsection{Screening current in an infinite superconductor with finite Cooper pair momentum $q$}

\begin{figure}[h] 
	\includegraphics[width= 1\columnwidth]{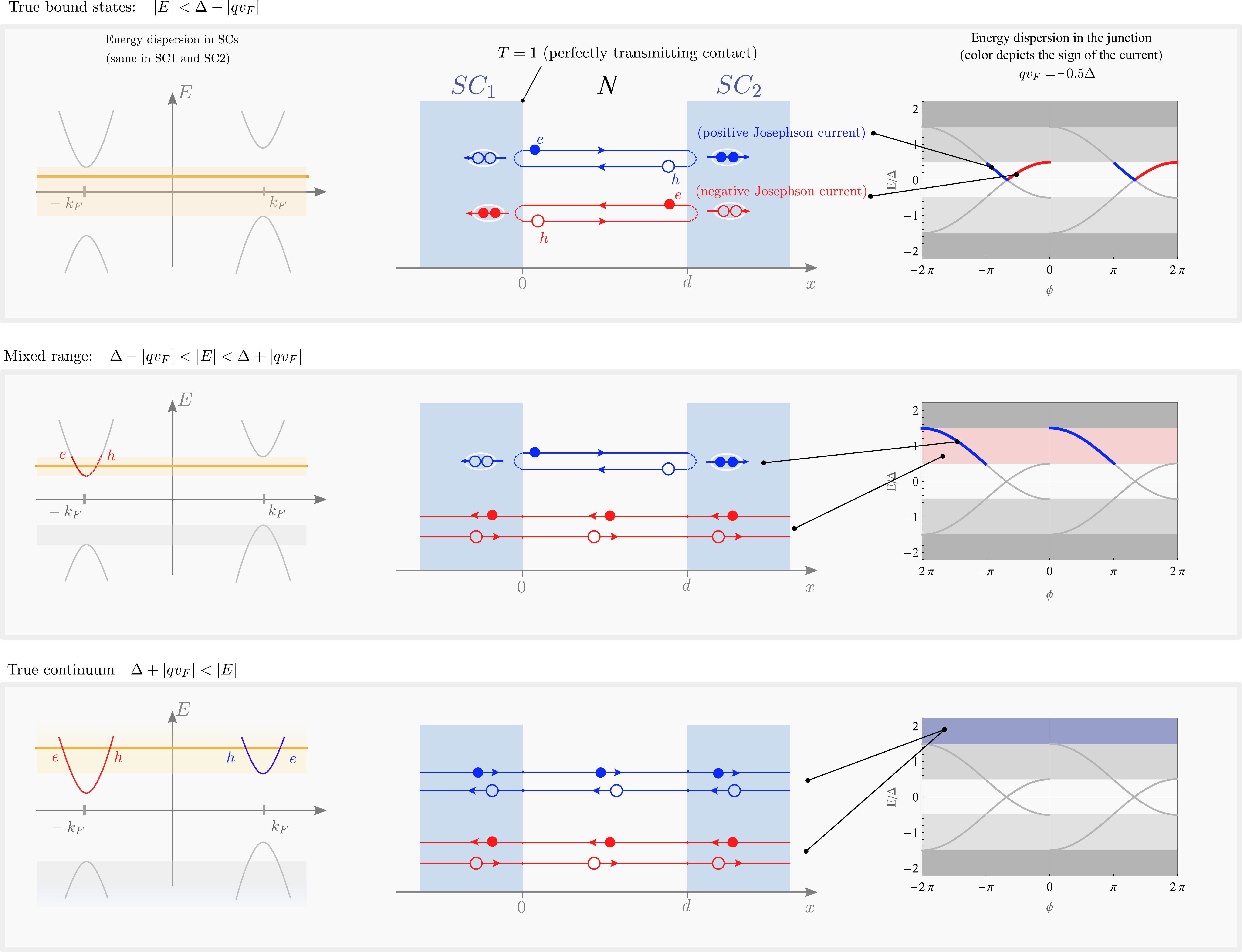}
	\caption{ An illustration showing the types of contributions to the Josephson current at different energy ranges in transparent junction. For simplicity, we only consider positive energy states because the spectrum is particle-hole symmetric. The left column shows the energy spectrum in an exteneded superconductor with finite-momentum pairing $\Delta (x)  = \Delta e^{2 i q x}$. The second and third columns show the schematics of the states in the junction and the energy spectrum in the junction, respectively.\\
	In the energy range $|E|<\Delta - |q|v_F $, there are truly bound Andreev states.
	In the presence of normal reflection, the bound state in the energy range $\Delta - |q|v_F < E < \Delta + q v_F$ in the middle panel becomes a quasi-bound state. At $|E| > \Delta + q v_F$, the truly continuum states exist, however, they are current-carrying, as we discuss in the text.
	} \label{ill}
\end{figure}

For the rest of the discussion, assume that $q$ is negative, which is the case in Fig. \ref{ill}. When the energy is in the range $\Delta - |q| v_F < |E|< \Delta + |q| v_F$, the left-moving states in the normal region correspond to a gapless energy range in both superconductors, as seen in Fig. \ref{ill}.

Let us compute the screening current that flows in an an infinite superconducting slab with $\Delta(x) = \Delta e^{2 i q x}$. For the energy range $\Delta - |q| v_F < |E|< \Delta + |q| v_F$ the current comes from left movers only:
\be
J_{1} = -e v_F \int_{\Delta - |q| v_F }^ {\Delta + |q| v_F}\nu_L(E) d  E =  -\frac{e}{   \pi \hbar} \int_{\Delta - |q| v_F }^ {\Delta + |q| v_F} \frac{E + |q| v_F}{\sqrt{(E + |q| v_F)^2 - \Delta^2} } d  E =   -\frac{2 e}{   \pi \hbar} \sqrt{|q|v_F (\Delta + |q|v_F)}
\ee
which, as we see, is zero when $q=0$. Analogously, the contribution from the true continuum states equals to a difference between the contributions:
\be
J_{2} = - e v_F\left ( \int_{\Delta + |q| v_F }^ {\infty}\nu_L(E) d  E  - \int_{\Delta + |q| v_F }^ {\infty}\nu_R(E) d  E  \right )=\frac{2 e}{   \pi \hbar} \left ( \sqrt{|q|v_F (\Delta + |q|v_F)} - |q| v_F \right)
\ee
Thus, the screening current is
\be \label{cont}
J_{scr}=J_{1} + J_{2}= \frac{2 eq v_F}{ \pi   \hbar }.
\ee
Which, as we see, equals \eqref{eq:I_cont}.

\clearpage
\section{Effects of normal reflection and disorder }

\subsection{Effect of normal reflection}

\begin{figure}[b] 
	\includegraphics[width= 0.9\columnwidth]{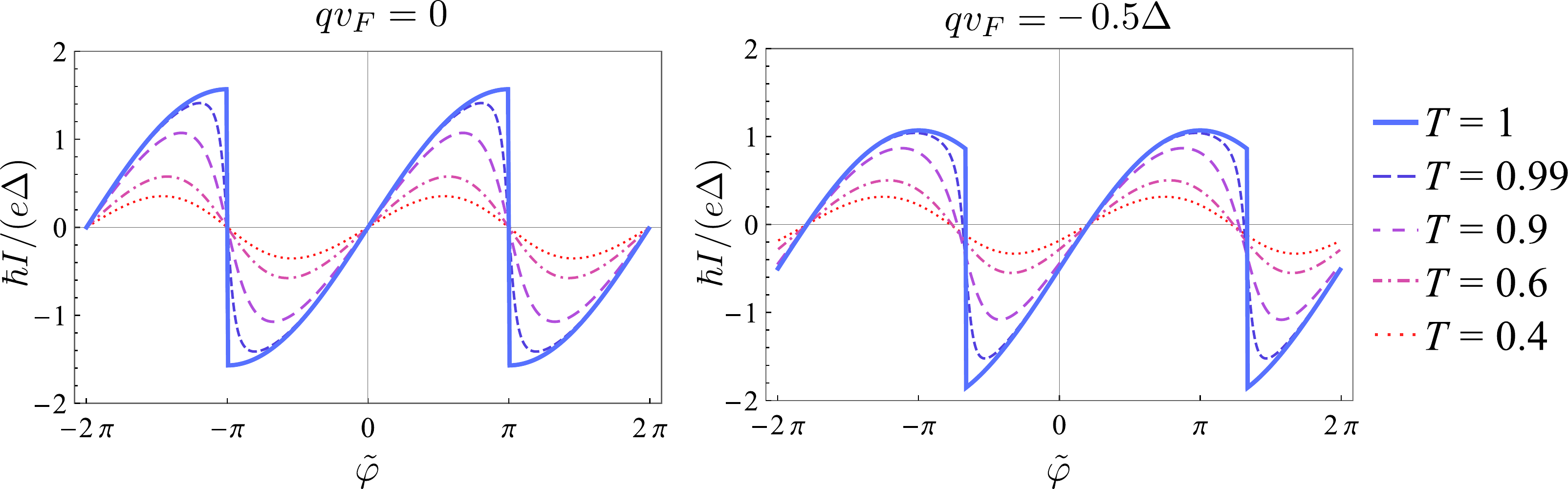}
	\caption{(a) Current-phase relations  at $q = 0$ and $q v_F = - 0.5 \Delta$ obtained by evaluating expression \eqref{supp_IT} at different values of barrier transparency.   
	} \label{fig:sm_current}
\end{figure}

Let us derive the expression determining the current-phase relation at finite barrier transparency, i.e. $T<1$. We use the notation $r_A^\pm = e^{i \gamma^\pm}$, where $\gamma^\pm(E) = \arccos{\frac{E \mp q v_F}{\Delta}}$ and further simplify \eqref{T}:
\be \label{supp:dos_T}
\det (1 - s_A s_N) \propto T \sin \left (\gamma^-(E) + \frac{\widetilde \varphi}{2}\right )\sin \left (\gamma^+(E) - \frac{\widetilde \varphi}{2}\right ) + (1-T) \left [ \sin^2 \left ( \frac{\gamma^-(E) + \gamma^+(E)}{2}\right)   \right] = 0
\ee
where, as a reminder $\widetilde \varphi = \varphi + 2 q d$. We use this expression for evaluation of the density of states as given in eq.~\eqref{Supp:nu} and follow the derivation of the Josephson current through eq.~\label{supp_I} with modified density of states according to eq.~\eqref{supp:dos_T}. We obtain the expression for the Josephson current through the barrier for a junction with finite transparency at zero temperature:
\be \label{supp_IT}
I(\varphi,q) = - \frac{e}{4 \pi \hbar} \int d \omega \Re \frac{ T \sin  \left ( \gamma^-(i \omega) - \gamma^+ (i \omega)  + \widetilde \varphi \right ) 
}{T \sin \left (\gamma^-(i \omega) + \frac{\widetilde \varphi}{2}\right )\sin \left (\gamma^+(i \omega) - \frac{\widetilde \varphi}{2}\right ) + (1-T) \left [ \sin^2 \left ( \frac{\gamma^-(i \omega) + \gamma^+(i \omega)}{2}\right)   \right]}
\ee
The current-phase relations computed from this expression at $T<1$ are shown in Fig.~\ref{fig:sm_current}. We see that result is similar to the one obtained from tight-binding calculation shown in Fig.~\ref{fig:SM_TB_spectrum}. 

For nearly transparent junction $T \approx 1$, as one can see from, the diode efficiency is non-analytical at $T\approx 1$. Thus, the current cannot be evaluated perturbatively in $(1-T)$ for nearly transparent junctions. 

At the same time, at small transparency $T \ll 1$, one can expand and obtain
\be \label{supp_IT}
\begin{split}
I(\varphi,q) = - \frac{e}{4 \pi \hbar} \int d \omega &\Re \left [ \frac{\sin (\varphi + \gamma^-(i \omega) - \gamma^+(i \omega))}{1- \cos (\gamma^-(i \omega) + \gamma^+ (i \omega))} T + \right. \\
&+ \left.\frac{\sin (\varphi + \gamma^-(i \omega) - \gamma^+(i \omega))}{4 \sin^2\left ( \frac{\gamma^-(i \omega) - \gamma^+(i \omega)}{2}\right) \left ( 1 - \cos(\varphi + \gamma^-(i \omega) - \gamma^+(i \omega))   \right)}  T^2  + O(T^3) \right]
\end{split}
\ee
The leading term $O(T)$ evidently leads to a symmetric current-phase relation and hence, no diode effect. The subleading terms lead to an asymmetric current-phase relation and emergence of the diode effect.  As we can see, the diode effect survives  at very small junction transparency $T \ll 1$.

% Fig.~\ref{fig:SMetaT}B shows the change in the slope of the diode efficiency at small $q$ with reflection coefficient $1-T$ and  illustrates the decrease of the effect due to normal reflections.

% \begin{figure}[t] 
% 	\includegraphics[width= 0.3\columnwidth]{derivative.pdf}
% 	\caption{Dependence of $d\eta/d q$ on the reflection coefficient $1-T$ at $q \rightarrow 0$. This provides a good measure for the change in the diode efficiency due to finite reflection coefficient at small and moderate values of $q$. 
% 	} \label{fig:SMetaT}
% \end{figure}

%Fig.}~\ref{fig:SM_disorder1} \hi{ 

\subsection{Effect of disorder}

We perform additional tight-binding simulations in order to gain some insight on the effect of disorder. 

 We use the parameters $\Delta = 2$, $t = 120$, $a = 1$ in our simulation and fix the length $N_N = 4$. In the figure labels below, we use $L$ and $N_S$ interchangeably.
\par
\begin{figure}[h] 
	\includegraphics[width= 0.6\columnwidth]{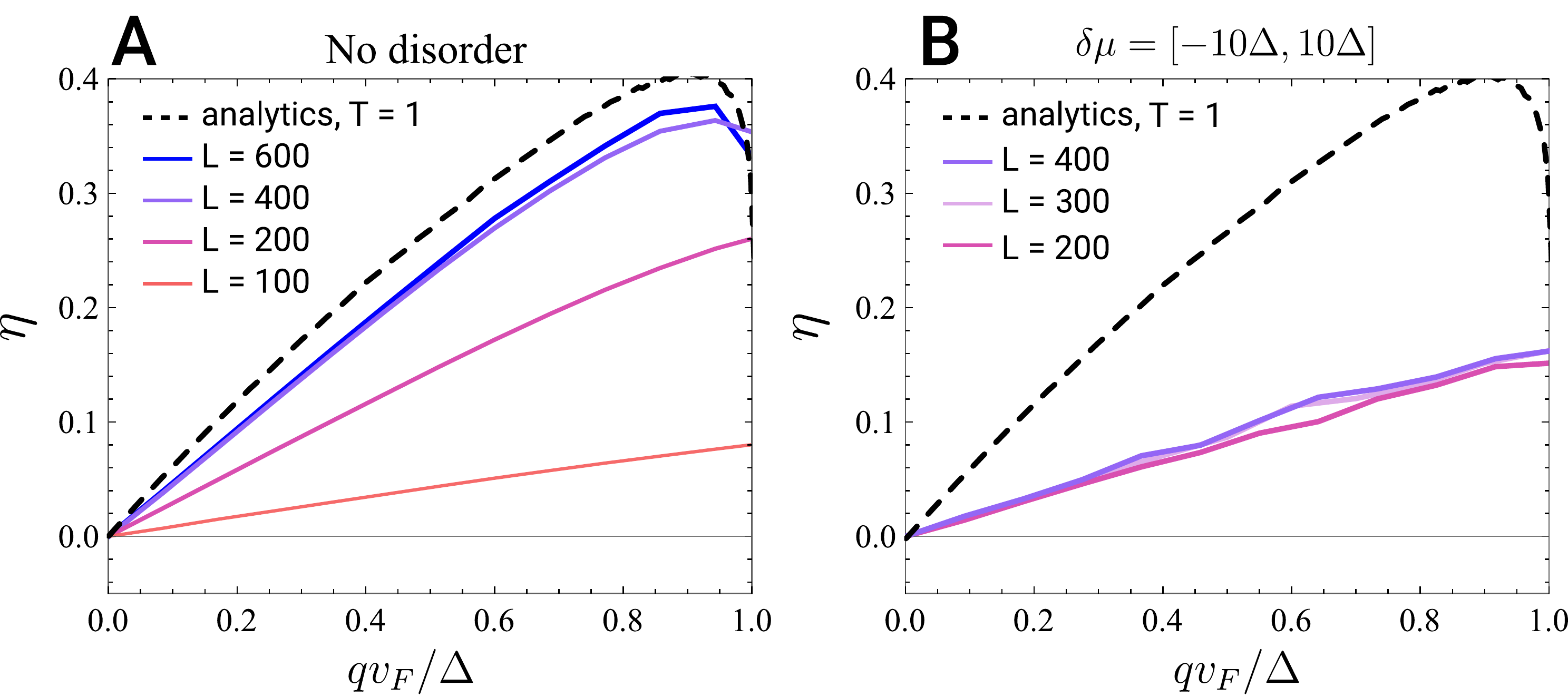}
	\caption{(A) Tight-binding calculations described in this section for a clean system for different length of the leads. The panel illustrates the sensitivity of the effect on the lead length that appears when $L/d<100$. (B) Diode efficiency  averaged over 150 realizations of chemical potential disorder uniformly sampled in the range$[-10 \Delta, 10 \Delta]$ for $L = 200,300$ and $400$. The black dashed line is the result obtained from eq.\eqref{supp_IT} at $T= 1$ for comparison. 
	} \label{fig:SM_disorder1}
\end{figure}

 In Fig.~\ref{fig:SM_disorder1}A  we show the result of the tight-binding simulation for a clean system at different length of the leads $N_S = L = 100,200,400$ and 600. We see that the diode efficiency at $L = 400$ and above converges to our analytical result shown by the black dashed line.   This panel demonstrates that the effect sensitively depends on the system length when $L/d < 100$, mainly because at this system length the level spacing of the levels in the 'continuum' (that turns into a set of discrete levels because of the finiteness of the system) becomes comparable to $\Delta$.
\par
Fig.~\ref{fig:SM_disorder1}B  shows the diode efficiency  averaged over 150 realizations of chemical potential disorder uniformly sampled in the range$[-10 \Delta, 10 \Delta]$ for $L = 200,300$ and $400$. As we see, at this value of the disorder the diode effect is still present, even though its magnitude is reduced by a factor of two. The effect also becomes system-length independent, which is because at this disorder strength the length $L$ becomes irrelevant in comparison to $\xi_{loc}\ll L$. This has to be contrasted with panel A.

\begin{figure}[h] 
	\includegraphics[width= 0.32\columnwidth]{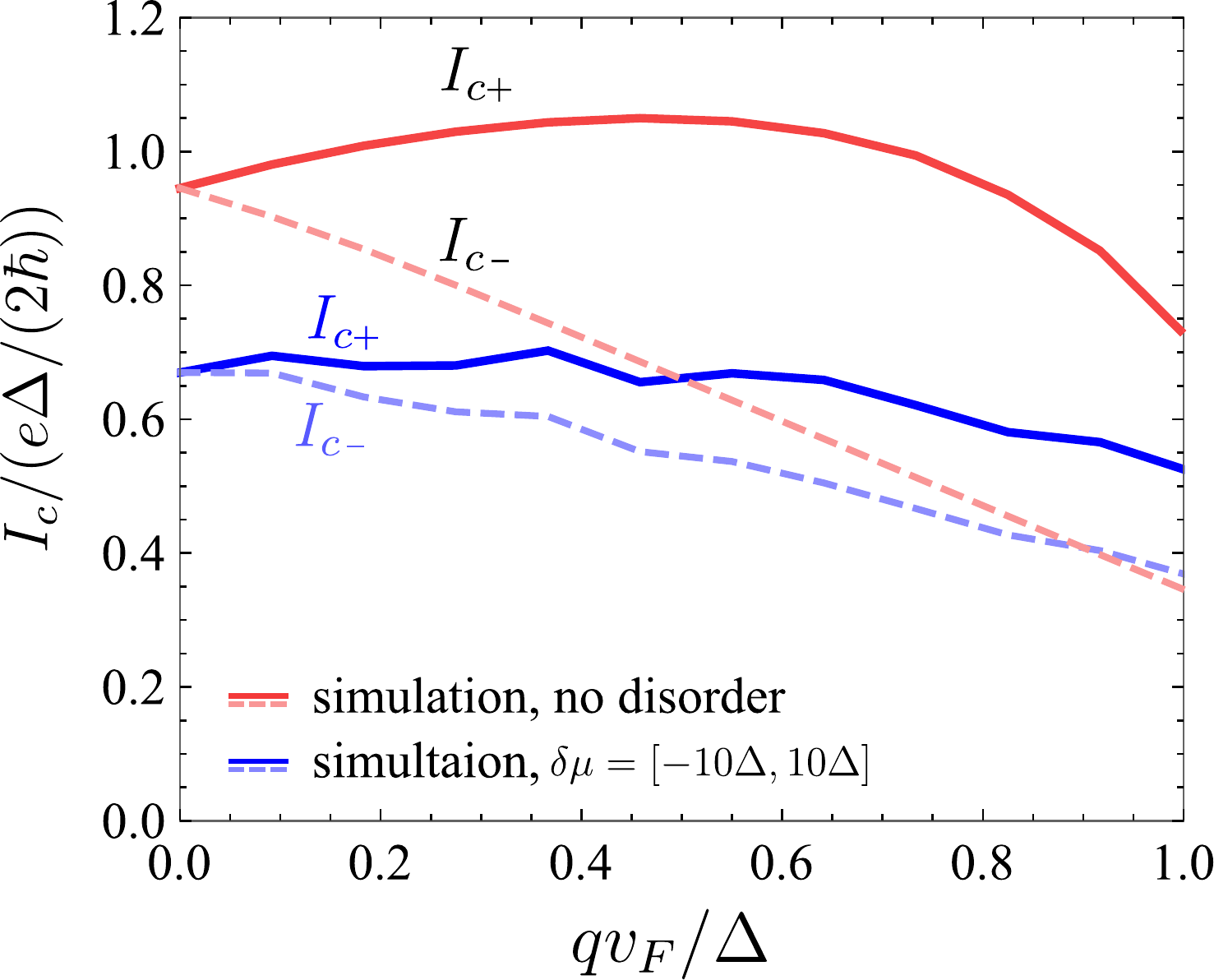}
	\caption{Dependence of the two critical currents $I_{c+}$ (solid lines) and $I_{c-}$ (dashed lines) on the Cooper pair momentum $q$. The red plots correspond to the clean system with $L = 400$, the blue ones are averaged over 150 realizations of chemical potential disorder uniformly sampled in the range$[-10 \Delta, 10 \Delta]$. %The black plots are obtained from eq.\eqref{supp_IT} at finite junction transparency $T= 1$ for comparison.
	} \label{fig:SM_disorder2}
\end{figure}

Fig.~\ref{fig:SM_disorder2}, shows the dependence of the critical currents on the Cooper pair momentum for L = 400. The red lines show the result for the clean system, the black one is the analytical result for the transparent junction for a reference. The blue plots correspond to the critical currents obtained after averaging over 150 realizations of chemical potential disorder uniformly sampled in the range$[-10 \Delta, 10 \Delta]$. We see that, even though the critical current reduces in value, it is still of the same order as the critical current in a clean system. 
\par
\begin{figure}[h] 
	\includegraphics[width= 0.6\columnwidth]{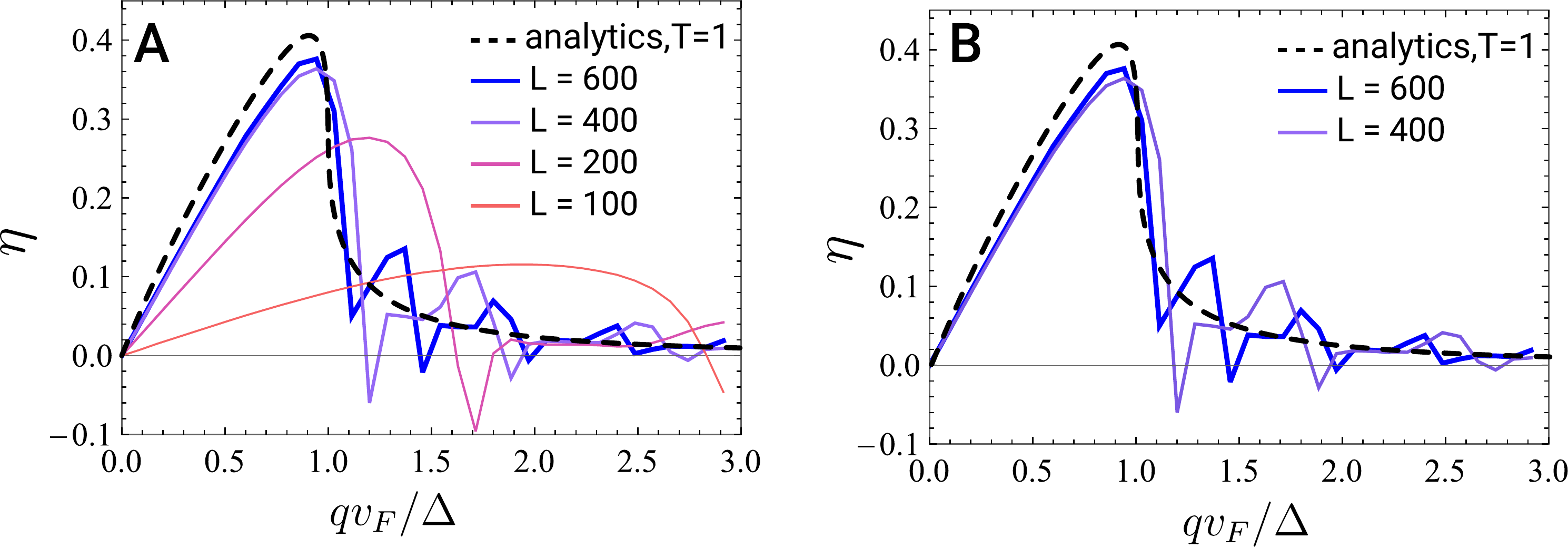}
	\caption{(A) Tight-binding calculations described in this section for a clean system for different length of the leads in a larger range of Cooper pair momenta. The black dashed line is the result obtained from eq.\eqref{supp_IT} at finite junction transparency $T= 1$ for comparison. 
	} \label{fig:SM_disorder3}
\end{figure}

 
In Fig.~\ref{fig:SM_disorder3} , the results of the calculations for the clean system are shown in a wider range of the Cooper pair momentum $q$. At $q> \Delta/v_F$, non-universal oscillations of the effect occur that depend on the system length.
\par

 We find that when $\xi_{loc}<L$, the sensitivity of the effect to the system length disappears entirely and its magnitude depends on the values of disorder and  the superconducting gap only. More importantly, the JDE occurs regardless of the ratio between the localization length  and the coherence length as long as $d, k_F^{-1} \ll \xi_{loc}$.  This demonstrates that the Josephson diode effect is universally robust. 

\end{document}